| | |
|---|---|
| **Title:** | **A unified-field theory of genome organization and gene regulation** |
| **Authors:** | Giuseppe Negro[1,2], Massimiliano Semeraro[2], Peter R Cook[3*] and Davide Marenduzzo[1*] |
| | *corresponding authors |
| **Address:** | [1]SUPA, School of Physics,<br>University of Edinburgh,<br>Peter Guthrie Tait Road,<br>Edinburgh, EH9 3FD,<br>UK.<br><br>[2]Dipartimento Interateneo di Fisica,<br>Universita` degli Studi di Bari and INFN,<br>Sezione di Bari,<br>Via Amendola 173, Bari, I-70126,<br>Italy.<br><br>[3]Sir William Dunn School of Pathology,<br>University of Oxford,<br>South Parks Road,<br>Oxford, OX1 3RE,<br>UK. |
| **Lead author:** | DM |
| **Correspondence:** | PRC    Telephone: (+44/0) 1865 275528<br>            Fax: (+44/0) 1865 275515<br>            E Mail: peter.cook@path.ox.ac.uk<br>DM    Telephone (+44/0) 1316 505289<br>            Fax (+44/0) 1316 505902<br>            E Mail dmarendu@ph.ed.ac.uk |
| **Key words:** | computer simulations, enhancer, expression quantitative trait locus (eQTL), silencer, super-enhancer, topologically-associating domain (TAD), A and B compartment, phase separation, polymer physics |

**Graphical abstract:**

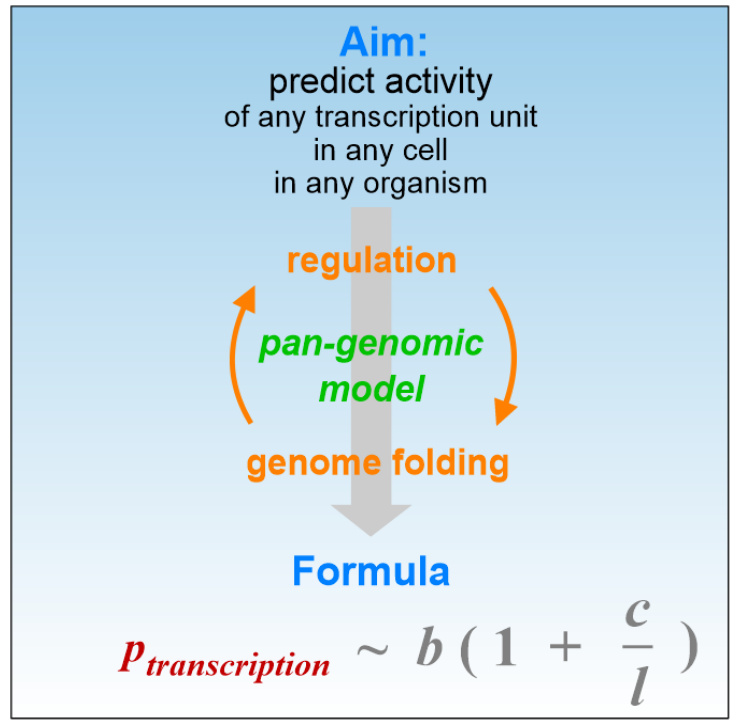

**ABSTRACT**

Our aim is to predict how often genic and non-genic promoters fire within a cell. We first review a parsimonious pan-genomic model for genome organization and gene regulation, where transcription rate is determined by proximity in 3D space of promoters to clusters containing appropriate factors and RNA polymerases – structures variously called transcription factories, hubs, and condensates. This model allows reconciliation of conflicting results indicating that regulatory mammalian networks are both simple (as over-expressing just 4 transcription factors switches cell state) and complex (as genome-wide association studies show phenotypes like cell type are determined by thousands of loci rarely encoding such factors). It also yields simple explanations of how mysterious motifs like quantitative trait loci, enhancers, and silencers work. We then present 3D polymer simulations, and a proximity formula based on our biological model that enables prediction of transcriptional activities of all promoters in three human cell types. This simple fitting-free formula contains just one variable (distance on the genetic map to the nearest active promoter), and we suggest it can be applied to any organism.

## Overview

Our ultimate aim in this work is to predict the rate of transcription of any promoter in a given cell type. To this end, we first briefly review some key principles underlying transcriptional initiation, and a biological model based on these principles. We then present results of 3D polymer model simulations based on this model, and derive a simple formula enabling prediction of the probability that a promoter might fire, which constitutes the main result of this work. We also compare results obtained using this formula with those obtained from simulations and experiments.

## The biophysics of transcriptional initiation: a mini-review

### Gene regulation: three possible mechanisms, and two elephants in the room

A gene might be transcribed differently in two different cells in the same organism for three main reasons (**Fig. S1A**). First, covalent structures of their DNA might differ. For example, cytosines at the 5' ends of many human genes become methylated during development, and bases in immunoglobulin genes are shuffled during B-cell maturation. However, most genes in most organisms in most cell states have similar covalent structures, so our focus on universal mechanisms means this possibility is not considered further. If covalent structures are identical, one gene might behave differently because it binds an activator or repressor, or it adopts a different 3D structure. We will argue interplay between the last two mechanisms self-organizes genomes (Misteli, 2020) and regulates activity.

Results of two powerful approaches dominate current thinking about gene regulation (**Fig. S1B**). Yamanaka's experiment points to a major role for transcription factors (TFs); over-expressing just four (Oct4, Sox2, c-Myc, and Klf4) converts mouse fibroblasts into a distinct cell type – induced pluripotent stem cells (iPSCs,; Takahashi & Yamanaka, 2016). Results from genome-wide association studies (GWAS) paint a very different picture. GWAS allows quantitative trait loci (QTLs) affecting any complex phenotype like the fibroblast or stem-cell state to be ranked; the sub-set influencing mRNA levels (and so transcription rates) are called expression QTLs (eQTLs) that are ordered according to levels of poly(A)$^+$ RNA determined by sequencing (RNA-seq; Stark *et al*., 2019). GWAS applied in man brought many surprises (Wray *et al*., 2018; Liu *et al*., 2019). First, QTLs are numerous and widely scattered, with both positive and negative effects that are individually modest. For example, in 2010 the 18 top-ranked QTLs affecting type 2 diabetes accounted for just 6% of expected heritability, and by 2015 more than 70% of the ~3,000 windows of 1 Mbp in the genome contained ≥1 QTL affecting schizophrenia. In both cases, thousands more QTLs are being identified as sample size and resolution improve. Second, QTLs rarely map to genes encoding TFs or even proteins; for example, most eQTLs are single-nucleotide polymorphisms (SNPs) in enhancers (Brynedal *et al*., 2017; The GTEx Consortium, 2017; Yao *et al*., 2017). Enhancers were originally defined as elements increasing transcription independently of orientation and position, but have since been defined in many other ways (Andersson *et al*., 2014; Schoenfelder & Fraser, 2019). We will adopt a definition used by FANTOM (Functional Annotation of the Mouse/Mammalian Genome) that sees them just as active non-genic promoters yielding enhancer RNAs (eRNAs) rather than mRNAs (Andersson *et al*., 2014). We also use the term "promoter" to describe a motif governing transcriptional initiation irrespective of whether the resulting transcript is a mRNA, eRNA, or other non-protein-coding RNA; consequently, there are many more non-genic promoters than genic ones in man (Andersson & Sandelin, 2020). When we use the term promoter, we will also assume this is active or potentially active in that cell type (as many are not active continuously and/or active in one cell type but inactive in another). Third, one gene is typically affected by many eQTLs/enhancers, and one eQTL/enhancer can target many genes that are often functionally related (Brynedal *et al*., 2017; The GTEx Consortium, 2017; Yao *et al*., 2017; Albert *et al*., 2018; Furlong & Levine, 2018; Schoenfelder & Fraser, 2019). Fourth, Hi-C (a high-throughput variant of 3C, chromosome conformation capture) shows eQTLs often physically contact their target genes (Javierre *et al*., 2016; Montefiori *et al*., 2018; Furlong & Levine, 2018), which points to contact (and so 3D conformation) playing a direct regulatory role.

Unfortunately, molecular mechanisms underlying eQTL and enhancer action are ill-understood, and corresponding models are complicated (Liu *et al*., 2019; Furlong & Levine, 2018; Schoenfelder & Fraser, 2019; Andersson & Sandelin, 2020). For example, applying the "omnigenic" model (Liu *et al*., 2019) to an eQTL would see a SNP altering activity of an enhancer with a consequential effect on transcription of the enhancer's target gene (which is rarely the ultimate eQTL target). Then, once the mRNA of the enhancer's target is translated, the resulting protein would rebalance regulatory networks in ways that depend on that target protein's role (e.g., by influencing signaling, sub-cellular localization, ATP levels, etc.). Finally, metabolic changes would percolate back in complex ways to modify transcription of the ultimate eQTL target. Note that no attempt is made in this model to explain why an eQTL might contact its ultimate target gene.

We summarize this section as follows. Results of two powerful experiments yield unreconciled findings: Yamanaka's result suggests regulatory networks are simple (just 4 TFs switch cell fate), but GWAS points to thousands of loci that rarely encode TFs tortuously determining phenotypes in complex post-transcriptional ways. We accept that regulation is multi-layered, but here our focus will be at the level of transcriptional initiation.

**Genome organization: DNA loops and the bridging-induced attraction**

DNA loops are major building blocks of chromosomes (Misteli, 2020). Those seen first in bacterial and human nucleoids, plus bacterial operons like *ara* and *lac*, were all anchored by the transcription machinery and had contour lengths of 20–200,000 bp (Stonington & Pettijohn, 1971; Jackson *et al.*, 1981, 1990; Schleif, 1992). Modern techniques like Hi-C (Rao *et al*. 2014), micro-C (Hsieh *et al*., 2020; Krietenstein *et al*., 2020; Zhang *et al.*, 2023), and GAM (genome architecture mapping; Beagrie *et al.*, 2017) confirm the presence of polymerases and TFs at anchor points. Note, however, that Hi-C misses many loops shorter than ~200,000 bp (Rowley *et al.*, 2020; Zhang *et al.*, 2021), and underestimates the presence of the transcription machinery at anchors (Beagrie *et al.*, 2021; Zhang *et al.*, 2023). Most anchors in archaea and plants are also transcribed (Le & Laub, 2016; Dong *et al.*, 2018; Cockram *et al.*, 2021; Takemata & Bell, 2021).

Loops are stabilized in four known ways. The classical mechanism involves dimerizing TFs; two TFs bind to nearby cognate sites on DNA to become trapped transiently in a local volume, and this increases the chances that they collide, dimerize, and anchor a loop (**Fig. 1A**; Rippe, 2001). Second, CTCF-cohesin complexes stabilize many mammalian loops (**Fig. S2A**; Rao *et al.*, 2014; Rowley & Corces, 2018). While CTCF is a TF, it is not encoded by plants or bacteria, and due to our focus on universal mechanisms we only discuss it in passing. In contrast, cohesin is conserved and anchors many long loops in addition to the ones mainly discussed here (Yatskevich *et al.*, 2019; Davidson and Peters, 2021), but it plays only a subtle role in human transcriptional regulation as knocking it down affects levels of only 23% expressed mRNAs and 1% unexpressed ones (Rao *et al.*, 2017). Third, many components of the transcription machinery also possess low-complexity disordered domains (including >80% mammalian TFs plus the catalytic sub-unit of RNA polymerase II; Liu *et al.*, 2006) that can coacervate into liquid droplets; then, phase separation involving components anchored at different sites can stabilize loops (**Fig. S2B**; Cramer, 2019; Rippe & Papantonis, 2021). Finally, the depletion attraction is another force that arises without energy input between large particles (e.g., DNA-bound polymerases) in a solution of smaller ones (e.g., nucleoplasmic proteins) that are sterically excluded from spaces between the larger ones; its strength is in the goldilocks zone – strong enough to overcome the cost of bending DNA, yet insufficient to ensure permanent contact (**Fig. S2C**; Marenduzzo *et al.*, 2006; Mitchison, 2019). With the exception of cohesin, all known looping mechanisms involve the transcription machinery.

We now describe another mechanism that clusters many loops. Consider the polymer model illustrated in **Figure 1B**, where red spheres (representing TFs complexed with RNA polymerase II, and which we call TF:pols) bind reversibly to pink beads (representing promoters and associated transcription units, TUs) scattered along a string of non-binding (or weakly-binding) blue beads (the rest of a chromosome; Brackley *et al.*, 2013; Brackley *et al.*, 2016). This model (and all subsequent ones that will be described) is based on few assumptions and is fitting free, in contrast to most models in the field. Strikingly, Brownian dynamics simulations show that bound TF:pols spontaneously cluster to generate many associated loops through an emergent process described as a "bridging-induced attraction" (bridges create an apparent attraction). Forming such bridges inevitably increases the local concentration of binding sites, and this triggers a positive feedback that operates without energy input to recruit more TF:pol complexes. When two different kinds of TF:pol complex (red and green) are simulated, distinct red and green clusters emerge (**Fig. 1C**). Additionally, if beads in a string representing human chromosome 19 in GM12878 cells are colored according to whether regions are active or inactive, loops, topologically-associating domains (TADs) and A/B compartments all appear without the need to invoke additional mechanisms (**Fig. S3**).

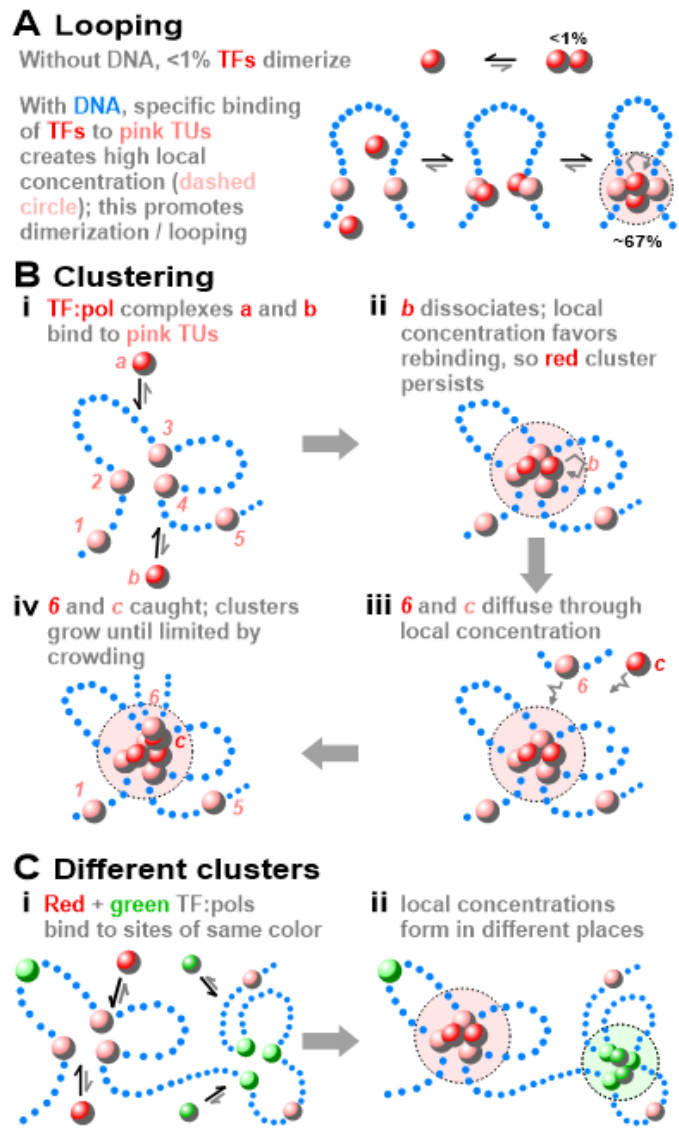

**Figure 1.** Local concentrations drive looping and clustering. Strings of non-binding blue dots: DNA. Pink spheres in the string: promoters/transcription units (TUs). Red and green spheres: TFs or TF:pols able to bind reversibly to ≥2 TUs simultaneously.
**A.** Classical model for TF-induced looping (from Rippe, 2001). TF concentration = 1 nM, dimerization equilibrium constant = $10^{-7}$ M (both typical values). Without DNA, <1% TFs dimerize. When TFs bind to promoters 10 kbp apart on DNA, they often collide to give a loop. This loop tends to persist as the local concentration (dashed circle) promotes TF rebinding (grey arrow); consequently ~67% TFs are now dimeric.
**B.** Clustering due to the bridging-induced attraction (from Brackley *et al.*, 2016). TF:pols *a*, *b*, and *c* bind reversibly to TU beads *1-5*. No other attractive forces between TF:pols or between TU beads are specified. (**i**) *a* and *b* will bind. (**ii**) When *b* dissociates, the local concentration of binding sites enhances its chances of rebinding (grey arrow). (**iii**) If *6* and *c* diffuse through the local concentration of factors, both are likely to be (**iv**) caught. Positive feedback (capture plus little loss) now leads to ~10 TF:pols/cluster until entropic crowding costs limit further growth.
**C.** Clusters of different types (from Brackley *et al.*, 2016). (**i**) Red TF:pols bind reversibly to pink TU beads, and green TF:pols to light-green ones. (**ii**) Positive feedback drives formation of red and green clusters in different places in 3D space, as pink and light-green TUs are in different places in 1D sequence space.

**Interplay between genome organization and transcription: a "pan-genomic" model**

Transcription can be linked to structure if one reasonably assumes that a TU is transcribed when it binds a TF:pol (Brackley *et al.*, 2021). Then, simulations yield patterns of transcriptional activity down chromosomes that are correlated in a statistically significant way with those obtained with GRO-seq (global run-on sequencing) – arguably the gold-standard reference (Jordán-Pla *et al.*, 2019; Stark *et al.*, 2019). Moreover, clusters resemble centers of activity seen experimentally and called phase-separated condensates/drops/pockets (Cramer, 2019; Palacio & Taatjes, 2021; Hilbert *et al.*, 2021), hubs (Boehning *et al.*, 2018; Monahan *et al.*, 2019; Misteli, 2020; Winick-Ng *et al.*, 2021; Ferrie *et al.*, 2022), clusters (Dotson *et al.*, 2022), super-enhancer (SE) clusters (Hnisz *et al.*, 2017), and transcription factories (Papantonis & Cook, 2013).

The bridging-induced attraction naturally yields complex interaction networks (Brackley *et al.*, 2021). To see why, consider the structure at the top of **Figure 2Ai** where TU bead *3* is bound to, and transcribed by, a TF:pol. When the TF:pol terminates, *3* is free to detach and diffuse across to bind to the right-hand cluster. Later, one (or both) of these clusters may disappear, and *3* may visit other clusters when they appear. Consequently, *3* contacts many other pink TUs over time to be co-transcribed with them. A regulatory network (Choy *et al.*, 2018; Liu *et al.*, 2018) can then be built (**Fig. 2Aii**) in which each TU is represented by a node, and pairs of nodes with positively-correlated activities are joined by black edges and anti-correlated ones by grey edges. Networks derived in this way from simulations of human chromosome 14 in HUVECs (human umbilical vein endothelial cells) are complex (**Fig. 2Bi**, left). Typically, black edges connect co-transcribed TUs in clusters that sequester TF:pols to reduce the likelihood that other clusters form elsewhere (giving grey edges). Strikingly, most nodes are highly connected (e.g., 63 of 67 TU beads in **Fig. 2Bi** are in the largest connected component, left), and small-world (most nodes are inter-connected by few edges). Moreover, pink and green TU beads in **Figure 1C** each form their own distinct small-world networks.

To model eQTL action, TF:pol binding to each of 39 TU beads in a toy string of 1,000 beads (each representing 3 kbp) was abrogated in turn (Brackley *et al.*, 2021). Each knock-out rewires the whole network in a distinct way; remarkably, activities of about half the other TUs both near and far away in sequence space change slightly, as the network retains its small-world character. Moreover, introducing non-binding "heterochromatin", binding "euchromatin", and permanent "loops" like those anchored by cohesin rewires networks in complex and difficult-to-predict ways. All these results fit comfortably with those seen by GWAS. Significantly, they result from co-transcriptional events that would act in addition to the post-transcriptional ones envisaged by the omnigenic model.

These results lead to a "pan-genomic" model that allows reconciliation of the two conflicting views of regulation (**Fig. S1B**). Thus far, TF:pol copy-number described in simulations reflects that found *in vivo*, so TU beads do not become saturated. However, increasing copy-number dramatically simplifies networks (**Fig. 2B**). We suggest this happens in Yamanaka's experiment: over-expression allows the system to escape from the complex small-world networks revealed by GWAS, so just 4 TUs can play their decisive roles (Brackley *et al.*, 2021).

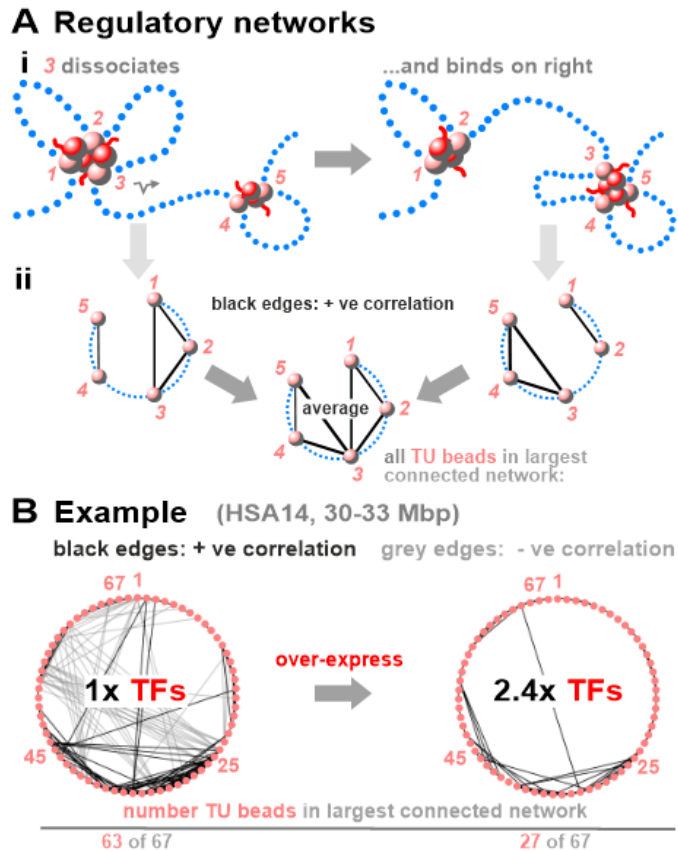

**Figure 2.** Reconciling results from Yamanaka and GWAS.
**A.** Regulatory networks (Brackley *et al.*, 2021). (**i**) During a simulation, *3* is in a cluster with *1* and *2* and "co-transcribed" with them as all lie close to TF:pols (large red spheres). *3* now detaches from the cluster and diffuses across to initiate in the right-hand cluster so its activity now positively correlates with those of *4* and *5*. Over time, TU beads start/stop being transcribed many times, and visit other clusters that appear/disappear. (**ii**) Contact and activity networks characterizing the two structures. Black edges indicate colocalization and positively-correlated activity, with the time average in the middle.
**B.** Effect of TF copy number on networks seen in simulations (time average given by 3 Mbp segment of HSA14 in HUVECs containing appropriately-positioned TU beads, each representing 3 kbp; for data, see SI). With the *in vivo* TF concentration ("1x TFs"), many TU beads contact other TU beads and have positively-correlated activities. However, TF:pols are in short supply, and binding to some TU beads necessarily decreases binding to others; this yields negatively-correlated activities indicated by grey edges. The network is small-world with 63 of the 67 TU beads in the largest connected network – consistent with GWAS results. Increasing the TF concentration 2.4x simplifies the network, which we argue allows Yamanaka's experiment to succeed.

We summarize thus far as follows. First, general arguments indicate that differential binding of TFs and/or folding underlie differential transcription. Second, loops are anchored by mechanisms usually involving the transcriptional machinery. Third, binding of this machinery triggers the bridging-induced attraction and clustering. Fourth, small-world regulatory networks then emerge spontaneously from the spatio-temporal dynamics of the system**.** Given these basic mechanisms, evolution has a choice: life forms can either spend energy to prevent such clusters and networks forming, or they can exploit these gifts of physics. We speculate the latter happened. Possibly, the depletion attraction clustered primordial polymerases transcribing DNA. Then, when cohesins evolved they stabilized additional loops, and – once TFs appeared – the bridging-induced attraction inevitably reinforced clustering and ensured that different clusters specialized in transcribing different gene sets, with disordered motifs strengthening clustering by enhancing phase-separation. From now on we call these emergent clusters "factories" – the first name for focal sites of transcription (Papantonis & Cook, 2013).

# Results

**A minimal model for transcriptional initiation**

In car factories, local concentrations of engines and tires facilitate efficient auto production. In transcription factories, analogous concentrations underlie efficient RNA production (e.g., the concentration of human RNA polymerase II in factories is ~1,000-fold higher than in the soluble pool). Consequently, the law of mass action ensures that essentially all transcription occurs in factories – as seen experimentally (Papantonis & Cook, 2013). Just as some car factories make Toyotas and others Teslas, the bridging-induced attraction drives formation of different transcription factories (**Fig. 1C**) that might make, for example, inflammatory or olfactory-receptor transcripts (Cook & Marenduzzo, 2018; Monahan *et al.*, 2019; Dotson *et al.*, 2022).

In all models for transcription, initiation frequency depends on how often promoters, polymerases, and TFs interact. In ours, there are two more key features. One concerns relative movement. The traditional model sees an active polymerase tracking down its template (**Fig. 3A**; **Fig. S4**). This idea stems from the perception by early biochemists of the relative size of a polymerase and its template – the smallest object (the enzyme) would move. But does it? We suggest it does not. Rather, it uses the energy released from the hydrolysis of nucleotide triphosphates to reel in its template – which has a much smaller cross-sectional area than what we now know to be a huge polymerizing complex (Cramer, 2019), and so the likeliest to move end on through the viscous milieu of a cell. Evidence for this has been reviewed (Papantonis & Cook, 2013; Cook & Marenduzzo, 2018), and we here discuss one example chosen because it is often cited as the best (and to our knowledge only) evidence for the traditional model – the iconic images of Miller spreads that show polymerases caught in the act of making RNA. In these static images, each transcript appears as an extended branch in a "Christmas tree" (Miller & Bakken, 1972) which it is assumed is made by a moving enzyme. However, a polymerase tracking along a helical path generates a transcript entwined about the template once for every 10 bp transcribed (**Fig. 3A**, bottom left), and not the extended and un-entwined branch seen in a spread. But if the template rotates as it is reeled in by a fixed polymerase, no entwinements result – as seen in the iconic images (**Fig. 3A** bottom right; **Fig. S5**, **Supplementary Movie 1**). Therefore, contrary to widespread belief, we suggest these spreads (and equivalent ones of lampbrush loops; **Fig. S5**) provide good evidence that active polymerases do not track.

The second key feature follows from the finding that most transcription occurs in factories/hubs (Papantonis & Cook, 2013): then, initiation frequency must depend on the distance in 3D space of a promoter to a factory rich in appropriate TFs. For example, in **Figure 3B**, promoter *f* is tethered closer to the factory than *g*; therefore, *f* is the more likely to visit the factory and initiate. Such distance effects provide simple explanations for how regulatory motifs work. Here, *e* acts as an enhancer of *f* because it tethers *f* close to an appropriate factory. **Figure S6** illustrates potential mechanisms based on this model for SEs, silencers, boundaries, insulators, eQTLs, and QTLs – as well as for other mysterious processes like transvection (Fukaya & Levine, 2017) and the pairing of meiotic chromosomes (Xu & Cook, 2008). Note that all these motifs are transcribed when active (Cook & Marenduzzo, 2018; Andersson & Sandelin, 2020). We also imagine that eQTLs and QTLs can act at the transcriptional level (and not just post-transcriptionally). Additionally, an active promoter is simultaneously any one of these motifs depending on which target gene is considered, and so it is no longer puzzling why these motifs are so similar at the molecular level (Papantonis & Cook, 2013; Andersson & Sandelin, 2020). This model also qualitatively explains why so many contacts seen by Hi-C, micro-C, and GAM involve transcribed regions (Rao *et al*. 2014; Hsieh *et al*., 2020; Krietenstein *et al*., 2020; Beagrie *et al*., 2017), with contact points moving as polymerases reel in their respective templates – which is impossible to explain without additional assumptions if polymerases track (**Fig. S4**). As expected, such contacts are sensitive to transcriptional inhibitors whereas contacts mediated solely by TFs or cohesin will be insensitive; then, it is also no longer puzzling (van Steensel & Furlong, 2019) that these inhibitors eliminate some loops but not others.

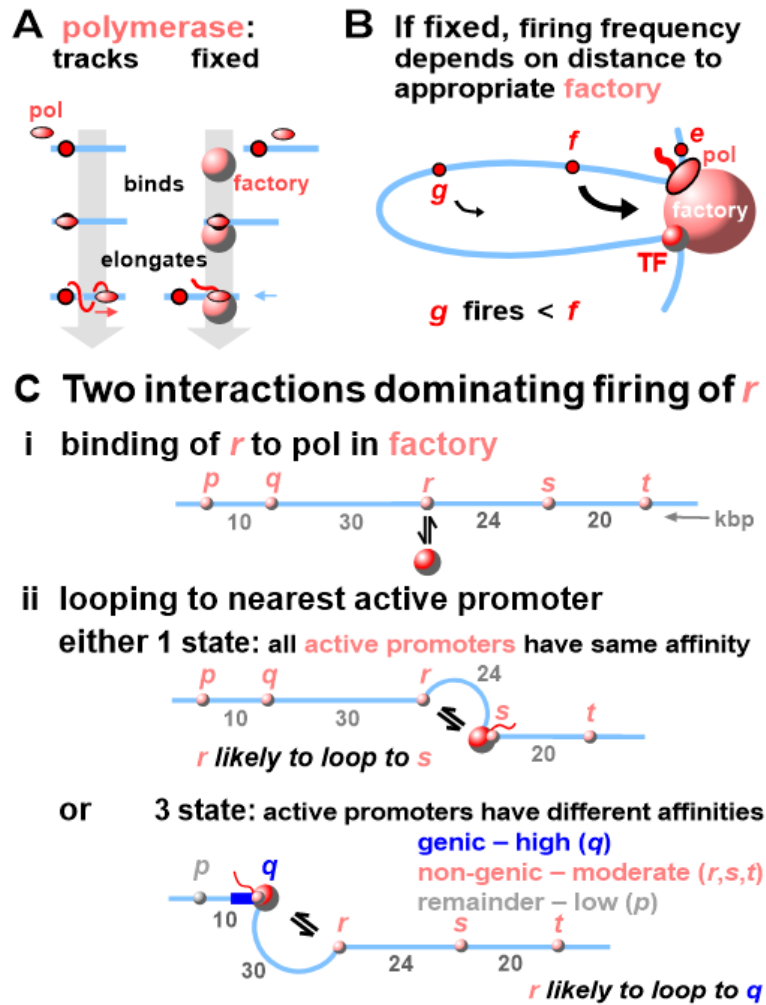

**Figure 3.** Models for transcription and gene regulation.
**A.** Transcription cycles. The conventional model (left) has a polymerase (pol) binding to a promoter (red circle), and tracking (pink arrow) down the template; however, the resulting transcript is inevitably entwined about the helical template once for every 10 bp transcribed, and there is no known mechanism for untwining it. The alternative (right) has a polymerase binding and reeling in (blue arrow) the template as it makes an un-entwined transcript (screwing a bolt through a fixed nut provides an analogy).
**B.** Initiation frequency depends on distance in 3D space to a factory. Here, $f$ is more likely to diffuse to the factory and fire than $g$. $e$ also acts as an enhancer of $f$ activity by tethering $f$ close to the factory.
**C.** Feynman-like diagrams depicting two interactions dominating firing of promoter $r$. Promoters: $p$-$t$. Factory: large red sphere. (**i**) Binding of $r$ to a factory. (**ii**) Forming a loop with the nearest active promoter on the genetic map. In the 1-state model, all active promoters have identical affinities for the factory. When the promoter nearest to $r$ is transcribed (i.e., $s$), $r$ becomes tethered close to a factory and so is likely to be transcribed. In the 3-state model, affinity of active genic promoter $q$ > active non-genic ones ($r$, $s$, $t$) > the "other" promoter ($p$). The 2$^{nd}$ Feinman diagram now involves the loop to the nearest genic promoter (and never to a non-genic or other one). Consequently, $r$ is often tethered close to a factory transcribing $q$ – and so often visits this factory to be transcribed.

**A simple looping formula for predicting firing frequency**

     Deriving formulae that facilitate prediction of gene activity has a long and continuing history (Davidson, 2010; Payne & Wagner, 2015; Sorrells & Johnson, 2015; Fulco *et al.*, 2019; Xiao *et al.*, 2020; Avsec *et al.*, 2021; Zuin *et al.*, 2022). For example, the deep-learning model "Enformer" shows great promise; it uses DNA sequence as input and is trained on a wide range of data sets that include DNase hypersensitive sites (DHSs), TF binding sites, and histone marks (Avsec *et al.*, 2021). In contrast to Enformer's top-down approach, here

we use a bottom-up one based on the model summarized in **Figure 3B**. In other words, our approach will be training-free and will not use any contact data as input.

Against the backdrop provided by GWAS that firing of any promoter is influenced by a myriad of eQTLs, we apply a strategy used by particle physicists who represent interactions dominating complex outcomes with a few Feynman diagrams. Consider potentially-active promoters *p-t* on a small segment of a chromosome, each with the same affinity for a factory; this will lead to what we will call the 1-state model. We suggest two Feynman-like diagrams depict interactions dominating the probability that *r* is transcribed ($p_{trans}$): the binding of *r* to a TF:pol in a factory (**Fig. 3Ci**), and the stabilization of an *r-s* loop (*s* being the nearest promoter to *r* on the genetic map, and so likely to be the closest potential tethering point; **Fig. 3Cii**). Then, the two terms in this approximate formula capture interactions in these two Feynman diagrams (see **SI** for a more detailed derivation):

$$p_{trans} \sim b\left(1 + \frac{c}{l}\right)$$

Here, *b* plus *c* are two positive constants that include contributions from TF:pol concentration, promoter number, affinity of TF:pols for promoters, plus a looping probability determined using the relevant polymer model – that of a fractal globule (Lieberman-Aiden *et al.*, 2009). The only variable, *l*, is the genomic distance in base pairs to the nearest active promoter. Note that the two Feynman diagrams depicting unlooped (**Fig. 3Ci**) and looped structures (**Fig. 3Cii**, 1-state) have equal weights in the case shown where *q* and *r* are separated by 30 kbp; however, looped:unlooped diagrams have weights of 10:1, 1:1 and 1:10 for loops of 8.6, 86, and 860 kbp, respectively.

Thus far, we have assumed all promoters have identical affinities. However, GRO-seq signals at active human genes turn out to be higher than those at non-genic TUs (Andersson & Sandelin, 2020) – possibly because TF:pols have a higher affinity for genic promoters. Therefore, we incorporate such a difference into constant *b*, and illustrate this using human chromosome 14 (HSA14) in HUVECs. We first identify all DHSs as they are such excellent markers for active promoters (Meuleman *et al.*, 2020), but others like ATAC-seq sites could be used instead. Next, these active promoters are divided into genic and non-genic ones (using the chromatin HMM browser track; Ernst *et al.*, 2011). Any remaining DHSs/promoters that have not yet been included are classified as "other". Values of constant *b* for these three states (i.e., "active genic" – $b_g$, "active non-genic" – $b_{ng}$, and "other" – $b_o$) are now weighted to reflect GRO-seq signals given by each type of promoter on this chromosome (data from Niskanen *et al.*, 2017). Additionally, when considering any kind of promoter, the 2[nd] Feynman-like diagram (and term in the formula) now involves a loop to another genic promoter (and never to a non-genic or other one). Consequently, the loop between *r* and genic *q* replaces the *r-s* one as the 2[nd] diagram in **Fig. 3Cii**, 3-state. We expect this 3-state model to perform better than the 1-state one in man, where active non-genic promoters so outnumber genic ones. Worked examples of the application of formulae to promoters *p-t* in **Figure 3C** are provided in **Supplementary Information** to highlight contributions of the two diagrams. Both the 1- and 3-state formulae are "looping formulae" as they capture spatial effects by looping to the most-proximal promoter.

**Testing performance of looping formulae: comparison with GRO-seq data**

We first test the performance of each formula on HSA14 in HUVECs. We identify all promoters active on the chromosome (using DHS data), determine $p_{trans}$ for each one (for the 3-state formula using values of constant *b* determined as above), and rank $p_{trans}$ values from high to low (**Fig. 4A**). Remarkably, patterns of activity given by the 3-state formula in two typical chromosomal segments better reflect those seen by our gold standard (GRO-seq) compared to poly(A)[+] RNA-seq – the most widely-used approach (**Fig. 4B**). Heatmaps show that $p_{trans}$ values obtained with the 3-state formula broadly match those from GRO-seq across the whole activity range from the least- to most-active (**Fig. 4C**).

We note that the value of the Spearman correlation between ranks predicted by the 3-state formula and those found by GRO-seq is relatively insensitive to exact values of *b* and *c* (**Fig. S7A**). For instance, for *c* = 86 kbp, any choice of $b = b_g, b_{ng}, 1$ for promoters producing mRNA, eRNA and neither, with $5 \leq b_g \leq 40$, $1 \leq b_{ng} \leq 8$, leads to a Spearman correlation larger than the one between GRO-seq and RNA-seq (**Fig. S7Ai,ii**). Additionally, for values of *b* used in **Figure 5B**, varying *c* between 3 and 300 kbp only changes the Spearman correlation from 0.66 to 0.62 (**Fig. S7Aiii**; **Fig. S7B** gives inter-chromosomal variations in values of $b_g$ and $b_{ng}$). Therefore, when applying the 3-state formula to cell types for which GRO-seq (or equivalent) data is unavailable, we expect that weightings for *b* applied in **Fig. 5B** can be used.

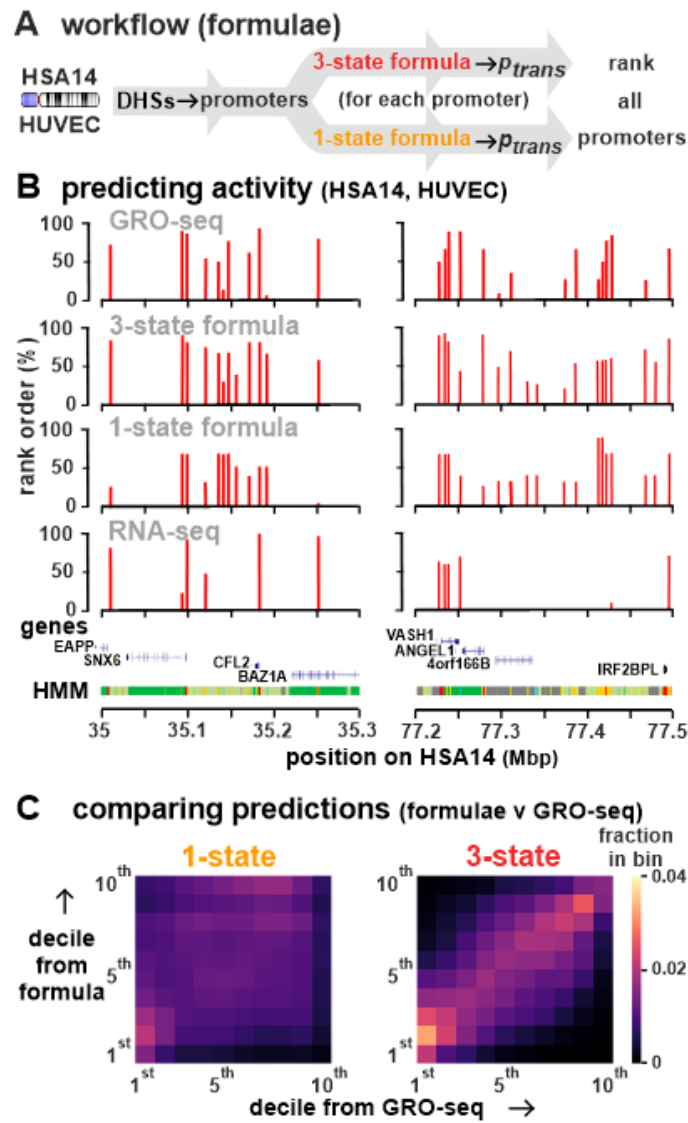

**Figure 4.** Testing the performance of looping formulae using HSA14 in HUVECs relative to results obtained with GRO-seq and RNA-seq.
**A.** Workflow. All active promoters are identified using DHSs data (from ENCODE), their $p_{trans}$ values calculated (using $c$ = 86 kbp for both formulae, and additionally $b_g$, $b_{ng}$, and $b_o$ = 13.1, 3.3, and 1, respectively, for the 3-state one), and rank orders of promoters determined. 2,226 identical promoters in the 1-state case are split into 344 genic, 938 non-genic, and 944 other in the 3-state case.
**B.** Formulae yield activity patterns in two typical regions of the chromosome similar to those obtained by GRO-seq (Niskanen *et al.*, 2017). Results are coarse grained into 3 kbp regions to allow comparison between data sets. The UCSC gene track excluding splice variants and non-coding genes (Hsu *et al.*, 2006), plus the chromatin HMM track (Ernst *et al.*, 2011), are included for reference (bottom). Profiles for RNA-seq miss poly(A)⁻ RNAs, but include stable mRNAs; they are chosen as comparators as they are so widely used to assess transcriptional activity, and to define eQTLs. However, it is nascent RNA levels that are of prime interest here, and not steady-state ones measured by poly(A)⁺ RNA-seq.
**C.** Activities predicted with the 3-state formula broadly match those from GRO-seq across the activity range. Ranked promoters are binned into deciles; bin color in the heat map reflects the fraction of all promoters found in a bin. Ten white squares on the diagonal from bottom left to top right (each with a fraction of 0.1) would represent a perfect match.

**Predicting transcription genome-wide in different cell types**
    We next extend this approach to all chromosomes in HUVECs, and to two other human cells for which GRO-seq data is available – lymphoblastoid GM12878, and embryonic stem-cell H1. Active promoters are

identified using DHSs, $p_{trans}$ values calculated, and Spearman correlations between computed rank orders compared with those determined using our gold-standard, GRO-seq (a value of 1 indicates a perfect match; **Fig. 5A**). With all cells, our simplistic 1-state formula provides a significant correlation usually higher than that expected by chance (**Fig. 5Bi-iii**; compare yellow bars with white circles). Remarkably, the 3-state formula slightly out-performs RNA-seq in HUVECs on all but HSA 3 and 21 (**Fig. 5Bi**, compare red bars with blue lines). With all cells, the 3-state formula again gives high Spearman correlations around 0.6 using values for $b_g$, $b_{ng}$, and $b_o$ obtained with HUVECs; clearly, these values prove to be transferable between cell types. However, now RNA-seq out-performs the 3-state formula with the other cells (**Fig. 5Bii,iii**; compare red bars with blue lines). When comparing the 1- and 3-state formulae, the 1-state model yields higher inter-chromosomal variation (e.g., gene-rich chromosome 19 gives low correlations) than the 3-state one.

In **Figure 5** we use values of $b_g$ and $b_{ng}$ obtained from GRO-seq on HSA14 and apply them to all chromosomes. As average ratios between GRO-seq signals at genic and non-genic promoters vary between chromosomes (**Fig. S7B**), we also tested values of $b_g$ and $b_{ng}$ that are cell- and chromosome-specific. Effects are marginal, and Spearman correlations between results from GRO-seq remain around 0.6 (**Fig. S8**). These results point to the 3-state formula providing an excellent and facile estimate of the transcriptional activity of all TUs (both genic and genic) in different cell types. Moreover, it can be even be applied to cells for which no GRO-seq (or equivalent) data is available to generate weightings for $b_g$ and $b_{ng}$ – as values derived from HUVECs prove to be transferable to other cell types (compare **Fig. 5Bi** with **Fig. 5Bii** and **iii**).

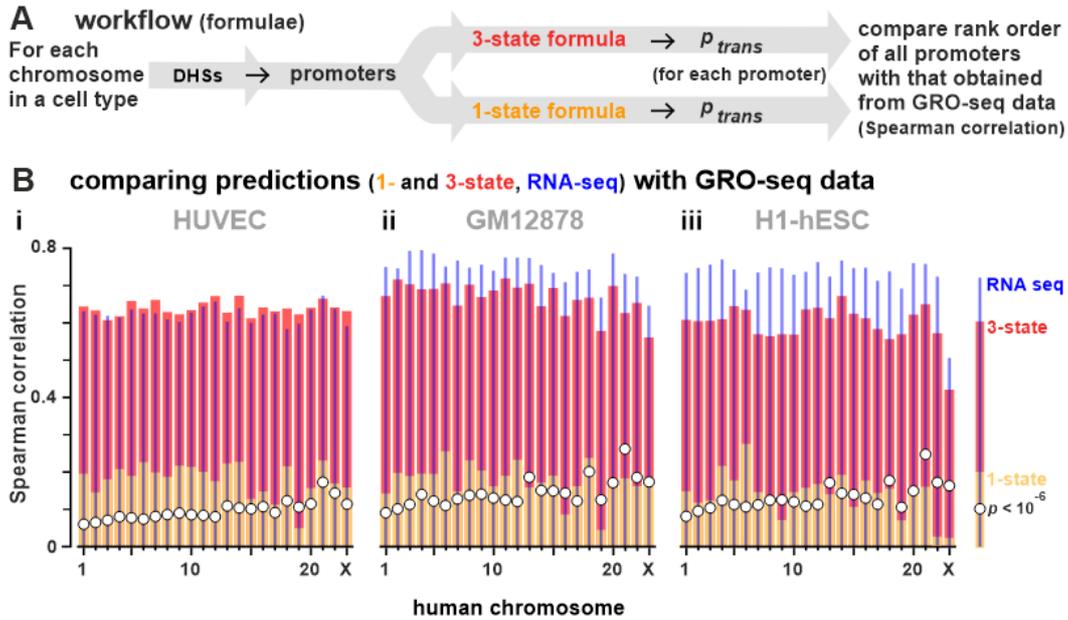

**Figure 5.** Predicting firing rates genome-wide.
**A.** Workflow. For each chromosome, all active promoters are identified using DHSs found in that cell type, their $p_{trans}$ values calculated and ranked from high to low, and Spearman correlations between the rank order and that from GRO-seq determined.
**B.** Spearman correlations obtained by comparing rank orders (for each chromosome in 3 human cell types) of firing probabilities determined using the formulae (and RNA-seq) with those from GRO-seq. For the formula, $c = 86$ kbp. For the 3-state formula, $b_g = 13.1$, $b_{ng} = 3.3$, and $b_o = 1$ (values determined using GRO-seq data for HSA14 in HUVECs are applied to all cell types). 1-state: yellow bars. 3-state: red bars. The p value computed measures the likeliness that correlations are obtained by chance.
(**i**) HUVEC. The 3-state formula and RNA-seq yield roughly similar correlations.
(**ii**, **iii**) GM12878 and H1-hESC. Values for $b_g$, $b_{ng}$, and $b_o$ are those for HUVECs and prove to be transferable to the two other cell types; however, RNA-seq now outperforms the 3-state formula.

**Testing performance of looping formulae: comparison with 3D simulation data**

We have seen that firing rates can be obtained from both simulations and formulae, and then validated by comparison with GRO-seq data. To provide additional validation (as well as a sanity check) we now complete the loop by comparing firing rates determined using formulae with those obtained from sets of simulations that mirror the 1- and 3-state conditions (**Fig. 6A**). In these Brownian-dynamics simulations, HSA14 in HUVECs is represented as a bead-and-spring polymer (a string of 35,784 beads, each of 30 nm diameter corresponding to 3 kbp) that is confined within an ellipsoid of appropriate volume (as individual chromosome territories are often ellipsoidal; Brackley *et al.*, 2021). Beads containing DHSs are identified as ones containing promoters, and those representing open chromatin by the presence of H3K27Ac histone-modifications. Simulations performed previously (Brackley *et al.*, 2021) provide 1-state data; in these, TF:pols were attracted to promoter-containing beads via a Lennard-Jones potential of 7.1 $k_BT$ (interaction range equal to 1.8 times bead diameter), and to beads marking acetylated regions more weakly (potential of 2.7 $k_BT$; Brackley *et al.*, 2021) – with all other beads being non-binding. For a set of 64 new 3-state simulations, promoter-containing beads are sub-divided into three types with different attractive potentials (i.e., genic – 7.1 $k_BT$, non-genic – 4.4 $k_BT$, and others – 3.5 $k_BT$), with the weakly- and non-binding beads being as for 1-state simulations.

**Figure 6B** illustrates a snapshot of a simulation volume at steady state in a 3-state run. As expected, clusters containing TF:pols bound mainly to genic and non-genic promoters spontaneously emerge. While these clusters occur throughout the volume, most (i.e., 88.5 ± 0.5%, where the error is standard error of the mean) are found in the outer 50% of the ellipsoidal volume (**Fig. 6**). This striking difference is in accord with results of "intron seqFISH" showing that most nascent human transcripts are found close to the surface of chromosome territories (Shah *et al.*, 2018). Such a peripheral location is also consistent with the greater numbers of *trans* versus *cis* chromosomal contacts seen by ChIA-PET after pulling down pol II (Li *et al*., 2012), and by GAM (Beagrie *et al.*, 2017). The mechanism driving clusters to the periphery is likely to be entropic: dispersing clusters with their high DNA densities as far away from each other as possible should reduce the free energy of the system. As before (Brackley *et al*., 2021), we determine $p_{trans}$ by measuring the amount of time a TF:pol spends bound to a bead. Both 1- and 3-state simulations lead to activity patterns down chromosomes that match those seen by GRO-seq (**Fig. 6C**). Both also yield Spearman correlations that correlate well with those given by formulae across the activity range (**Fig. 6D**). This provides further validation of the effectiveness of the looping formulae, and that formulae can replace 3D simulations when computational resources are limited, or if calculation time needs to be minimised.

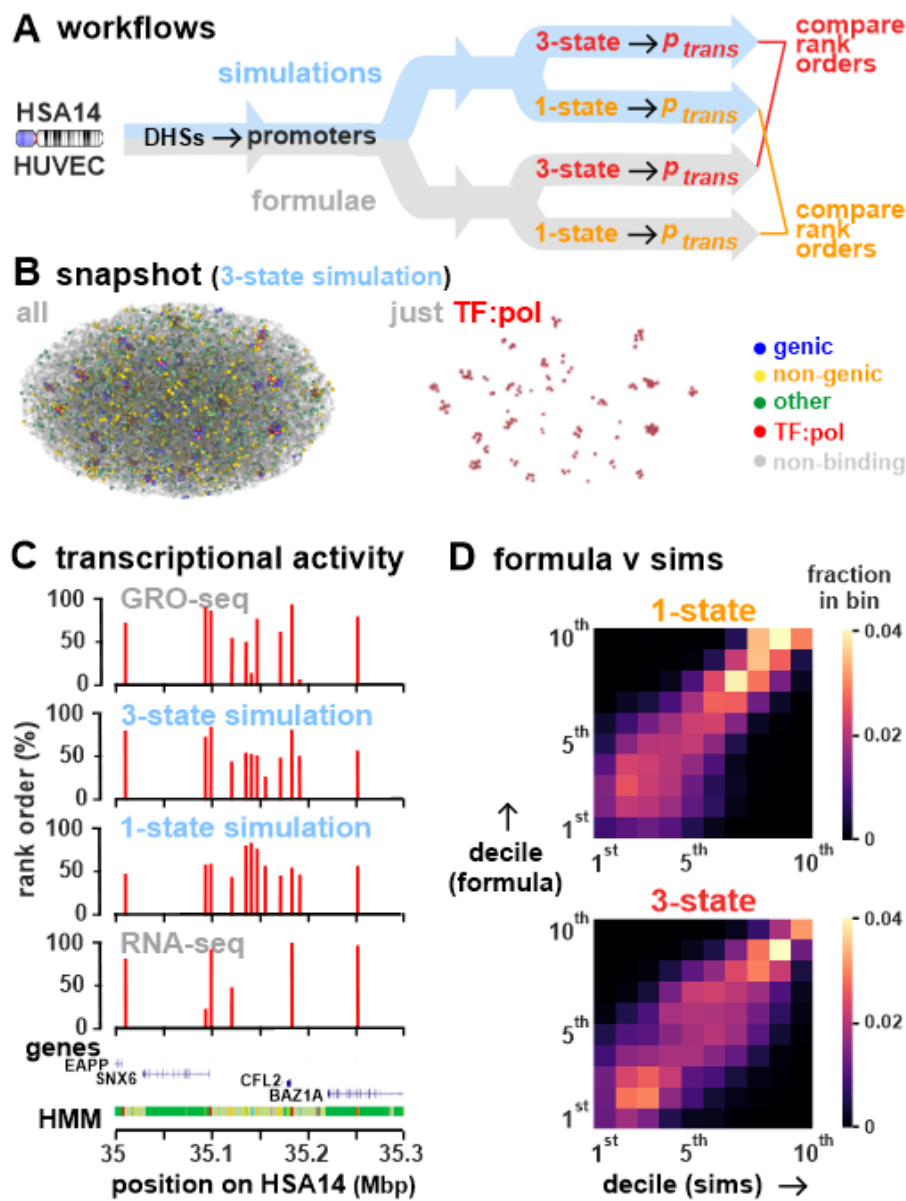

**Figure 6.** Comparing firing rates obtained using formulae and simulations.
**A.** Workflow. For HSA14 (HUVEC), all active promoters are identified using DHSs; then, using either simulations or formulae, $p_{trans}$ values are calculated, ranked from high to low, and rank orders compared.
**B.** Snapshot from a 3-state simulation (left – all beads, right – just TF:pol beads).
**C.** Simulations yield activity patterns in a typical chromosomal segment like to those obtained by GRO-seq. Data for 1-state simulations are from Brackley *et al*. (2021), and for 64 new 3-state simulations, see Supplementary Information. Coarse-graining is as in Figure 4B, from where GRO-seq, poly(A)+ RNA-seq, gene, and HMM tracks are reproduced for comparison. Spearman correlations and *p*-values with GRO-seq data are respectively: (i) 0.32 and <10$^{-6}$ (simulations, 1 state); (ii) 0.60 and <10$^{-6}$ (simulations, 3 states); (iii) 0.62 and <10$^{-6}$ (poly(A)+ RNA-seq). Another comparison can be done by binning rank data into deciles. Then, we count how many data fall in the same bin from GRO-seq and the other tracks, normalising this count by the number expected by chance: this ratio measures how good agreement with GRO-seq is. Ratios are: (i) 1.31 (1-state simulations); (ii) 1.82 (3-state simulations); (iii) 1.71 (poly(A)+ RNA-seq).
**D.** Activities predicted with the 3-state formula broadly match those from simulations (sims) across the activity range. As in Figure 4C, ranked promoters are binned into deciles, bin color in the heat map reflects the fraction of all promoters found in a bin, and 10 white squares on the diagonal from bottom-left to top-right (each with 0.1 counts) would represent a perfect match.

## Conclusion

Our aim was to predict how transcriptionally active a gene might be in any given cell under any given condition – against the background that gene regulation appears to be both simple (only 4 Yamanaka TFs switch cell fate) and complex (a myriad of eQTLs determine complex phenotypes, with few of these eQTLs encoding TFs; **Fig. S1B**). Therefore, we reviewed a parsimonious pan-genomic model for both the organization and regulation that accommodates these very different results (**Figs. 3B**). In this model, transcription rate depends mainly on the frequency with which a promoter visits a factory rich in appropriate TFs, and this enables us to propose simple mechanisms for how mysterious motifs like enhancers, silencers, and eQTLs all work (**Fig. 3B and S6**). Note that these factories are unlike car factories with stable architectures; instead, they are ephemeral, morphing as TF:pols bind/dissociate, and loops appear/disappear. Moreover, chromosome conformation is unlikely ever to be the same in two daughter cells in any tissue in our bodies simply because there are so many promoters able to visit so many appropriate factories, with any active promoter influencing the activity of all others to some degree (Brackley *et al.*, 2021). Then, the logic developed during evolution that is embedded in DNA sequence, TF concentration, and binding frequency acts through the bridging-induced attraction to organize small-world networks of clusters and loops that determine initiation rates.

Motivated by this pan-genomic model, and by focusing on just two configurations that we suggest dominate outcomes (i.e., binding to a factory, and forming the most-likely loop), we went on to develop two variants of a simple looping formula that enable prediction of the probability that any promoter (whether genic or non-genic) is transcribed. The 1-state formula treats all promoters identically, and requires as input only the number of bp to the next nearest promoter on the chromosome. The 3-state formula divides promoters into three groups that fire at different rates (i.e., genic, non-genic, plus ones not included in either of these two lists). When this 3-state formula is applied to 3 human cells (HUVECs, lymphoblasts, stem cells), it performs as well as RNA-seq and 3D polymer simulations in estimating the firing rates of all promoters across the whole activity range, and gives Spearman correlations of ~0.6 when rank orders of firing probabilities are compared with those obtained from GRO-seq (**Figs 4,5,6**). Consequently, these looping formulae have various general advantages. They are based on the physical forces we suggest drive genome conformation and promoter firing, they depend mainly on only one variable (distance in base-pairs to the nearest active promoter) and not on data-fitting and/or machine-learning used by other promising (but top-down) approaches (e.g., Ronquist *et al.*, 2017; Avsec *et al.*, 2021), they are easily extended (e.g., by adding more diagrams that include *trans* contacts and/or different TF:pol complexes binding with different affinities, and by varying weightings between different diagrams), and they are general in that they can be applied in any organism (however they can only be validated when appropriate GRO-seq or equivalent data is available; see **Supplementary Information**). At the same time, we stress that there are some assumptions in our derivation of the looping formulae, which ultimately limit their applicability. The main ones are the following. (i) We include only two diagrams, and do not include *trans* contacts. (ii) We do not include different types of TUs (for instance, binding to different types of transcription factors). (iii) In the 3-state formula, we only loop in the 2$^{nd}$ diagram to the nearest genic promoter (however far away it might be, and never to a non-genic or "other" one that might be closer). We anticipate that in the future these assumptions will be tuned to accommodate new data, and to add complexity so that a higher Spearman correlations with GRO-seq data can be achieved.

We hope our unified conceptual views of genome organization and gene regulation can be combined with others to eventually enable us to predict a gene's activity and alter it in any desired way. The result of Yamanaka's experiment coupled with the success of the "AlphaFold2" algorithm in solving the protein-folding problem (Jumper *et al.*, 2021; Tunyasuvunakool *et al.*, 2021) – which is analogous to the genome-folding problem outlined here – encourages us in our hopes. In **Supplementary Figure 9** we describe a roadmap that addresses a grand challenge – how one might switch the fate of any human cell in a desired way.

**Acknowledgements**

This work was supported by the European Research Council (CoG 648050, THREEDCELLPHYSICS; DM). We thank C. A. Brackley and M. Chiang for helpful discussions.

**Conflicting interests**

None. PRC was a founder of, and holds equity in, iotaSciences Ltd. – a company exploiting a microfluidic technology unrelated to the subject matter of this manuscript.

**Author Contributions**

GN, MS and DM performed the simulations and calculations; GN, MS, PRC and DM wrote the paper.

**SUPPLEMENTARY INFORMATION**

**A looping formula relating initiation rate to distance from an appropriate factory**

Consider a genomic segment with *N* TU beads that bind *n* TF:pol complexes; our formula will yield the probability that a given promoter in a TU, *i*, is transcribed – which we estimate as the probability that a TF:pol is bound to *i*. Many configurations contribute to this probability, but two do so significantly as they are the first steps in the firing pathway (depicted in **Fig. 3Cii** as Feynman-like diagrams): one where *i* binds to a polymerase in a factory, and the second where *i* is tethered by the nearest active promoter to that factory. As we consider only promoters that are active or potentially active in the cell type under consideration, the Boltzmann weight of the unlooped configuration is $n\, e^{\beta\epsilon}(1-p(l))$, where $\beta = 1/(k_B T)$ with $k_B$ the Boltzmann constant and $T$ the temperature, $\epsilon$ is the affinity between TF:pol and *i* (which is initially assumed to be same for all promoters for simplicity in the one-state model; **Fig. 3Cii**), $l$ is the distance on the genetic map in base pairs between *i* and the nearest promoter, and $p(l)$ is the probability that the segment is looped (anchored by *i* to its nearest promoter). The Boltzmann weight of the looped configuration is $n\, e^{2\beta\epsilon}p(l)$, because now there are two contacts between the TF:pol and a TU.

For a confined chromosome, the relevant polymer model is that of the fractal globule, for which the looping probability can be approximated as $p(l) \sim \frac{a}{l}$, with $a$ a suitable constant (Lieberman-Aiden *et al.*, 2009). The sums of the Boltzmann weights of looped and unlooped configurations (**Fig. 3Cii**) is then $n\, e^{\beta\epsilon}\left[1 + \frac{(e^{\beta\epsilon}-1)a}{l}\right]$, and the transcriptional probability is this sum divided by the sum of weights of all possible configurations (which we denote by $Z$, and whose explicit form we do not require here because we will only consider rankings of transcriptional activities, so that its value drops out of the analysis). Then, the probability that *i* is transcribed ($p_{trans}$) has the form:

$$p_{trans} \sim b\left(1 + \frac{c}{l}\right)$$

where $b$ plus $c$ are two positive constants ($b$ includes contributions from TF:pol concentration, promoter number, and the affinity of TF-pols for promoters; $c$ includes contributions from the affinity of TF-pols for promoters, and looping). For example, the unlooped configuration in **Figure 3Cii** gives a weight of $b$, and the looped one a weight of $\frac{bc}{l}$, where $l = 24$ (i.e., distance in kbp between *r* and *s*). Note that as ranks of transcriptional activities are compared in subsequent statistical analyses, the two constants $b$ and $c$ become irrelevant in this case (because they divide out and do not contribute to the relative rank).

This simple formula, which we refer to as the "looping formula" to highlight the fact that it captures the effect of 3D looping to proximal promoters, can be generalized to situations where different promoters bind TF:pol complexes with different affinities, and the approach that follows is motivated by the observation that active genes in GM12878 cells were known to yield 2- to 3-fold higher GRO-seq signals than active enhancers (Core *et al.*, 2014; Andersson & Sandelin, 2020). For instance, consider the case where a promoter (corresponding in our example to a DHS peak) produces either a mRNA, or an eRNA (or other non-genic RNA), or one that is in neither of these two lists. These three states are identified using the ChromHMM browser track (Ernst *et al.*, 2011) in which active genic promoters are marked by HMM state 1, active enhancers and other non-genic RNAs by HMM states 4 + 5, and some DHSs have none of these HMM states (these will constitute our "other" class). In the "3-state model", we subdivide constant $b$ into 3 types that apply to TUs that are "active genic" – $b_g$, "active non-genic" – $b_{ng}$, and "other" – $b_o$. Then, when considering any promoter *i* (whether it is genic, non-genic, or other), the loop in the 2$^{nd}$ Feynman diagram connects *i* to the nearest genic promoter (e.g., *r* connects to genic *q* in **Figure 3Cii**, 3-state), and constant $b$ now captures *i*'s state (i.e., it is higher for such a promoter producing a mRNA, intermediate for one producing an eRNA, and smaller for the "other" class). To reiterate, when considering any kind of promoter, we only loop in the 2$^{nd}$ diagram to another genic promoter (and never to a non-genic or other one).

We now provide worked examples where each formula is applied in turn to promoters *p-t* in **Figure 3C**. First, consider the 1-state formula. Constants $b$ and $c$ contain components capturing TF:pol and promoter concentrations, plus the affinity of TF-pols for promoters. As our aim is to compare relative values of $p_{trans}$ and determine a rank order, and as each of these components is common to every promoter, we note that both *b* and *c* are actually irrelevant in the 1d model (they will divide out). However, in what follows we include these constants in the calculation for clarity. As the average loop length is not known for the cells analyzed, we use the average length found in HeLa (i.e., 86 kbp; Papantonis & Cook, 2013) and set this equal to *c*. Then, for promoter $r$, $p_{trans} \sim b \left(1 + \frac{86}{24}\right) \simeq 4.58\, b$ (as $c = 86$ kbp and $l = 24$ kbp) and values for $p - t$ are 9.6 $b$, 9.6 $b$, 4.58 $b$, 5.3 $b$, 5.3 $b$ respectively; this gives a rank order (from high to low) of $p = q, s = t, r$. Now consider the 3-state formula. For promoter $r$, we note that the distance to the closes genic promoter is now $l = 24$ kbp, so that $p_{trans} \sim b_{ng} \left(1 + \frac{86}{30}\right) \simeq 3.87\, b_{ng}$. Values for $p - t$ are 9.6 $b_o$, $b_g$, 3.87 $b_{ng}$, 2.59 $b_{ng}$, 2.16 $b_{ng}$, respectively (note that the estimate of firing of $q$ includes only the first diagram as calculating the second requires knowing the location of the closest genic promoter which is not shown in the sketch). In **Figure 4B**, we use values for $b_g$, $b_{ng}$, and $b_o$ = 13.1, 3.3, and 1, respectively, which gives a rank order (from high to low) of $q$, $r$, $p$, $s$, $t$. These values of $b$ are based on observed average ratios between GRO-seq signals for mRNA, eRNA, and other DHS peaks in human chromosome 14.

Our theory can be developed to include additional topologies (e.g., *trans* contacts, as well as double loops as in **Fig. 1Bii**), and different types of TF:pol (e.g., to include red plus green TFs binding to pink and light-green promoters respectively, as in **Fig. 1Cii**). The latter is especially relevant when modeling different cell types (as in **Fig. S9**). In this case, looped Feynman diagrams would connect promoters of the same type/color. Note that simulations of TFs with 5 different colors each binding with a different affinity to 5 different kinds of TU bead leads to clusters that mainly contain TU beads of the same color (Brackley *et al.*, 2016). Consequently, this formula can be extended to tens of different TFs when analyzing complex mammalian cell states and addressing the grand challenge outlined in **Figure S9**.

Data for the "formula" tracks in **Figure 4B** were derived as described above. Note that the value of the Spearman correlation between ranks predicted by the 3-state formula and those found by GRO-seq is relatively insensitive to exact values of *b* and *c*. For instance, for values of *b* used in **Figure 4B** varying *c* between 3 and 300 kbp changes the Spearman correlation from 0.66 to 0.62 (**Fig. S7**). Additionally, for $c = 86$ kbp, any choice of $b = b_g, b_{ng}, 1$ for promoters producing mRNA, eRNA and neither, with $5 \leq b_g \leq 40$, $1 \leq b_{ng} \leq 8$, leads to a Spearman correlation larger than the one between GRO-seq and RNA-seq. Therefore, when applying the 3-state formula to cell types for which GRO-seq data (or equivalent ATAC-seq data) is not available, the exact weightings for *b* are unlikely to be crucial, and could be estimated based on our results or using data from close relatives for which data is available.

Data for the "simulations" tracks in **Figure S3B** were taken from Brackley *et al.* (Brackley *et al.*, 2021) for the 1-state case. For the 3-state case, we ran 64 analogous simulations where TF:pols bind to TU beads with a Lennard-Jones potential with interaction range (as before) equal to 1.8 times the size of a chromatin bead, and depth equal to 7.1 $k_BT$ (genic promoters), 4.4 $k_BT$ (non-genic) and 3.5 $k_BT$ (all others). Interactions between TF:pols and non-TU chromatin was as in Brackley *et al.* (Brackley *et al.*, 2021).

Note that the formula can be applied to any organism for which chromosomal positions of all promoters active in the cell type under study are available (which can be obtained, for example, using GRO-seq, ATAC-seq, or inspection of appropriate histone marks). However, a critical test of how accurately our formula (or any other approach) enables prediction of activity requires lists of relative activities of all transcription units in a genome (i.e., both genic and non-genic). Whilst lists of relative numbers of steady-state RNAs are widely available, those of relative numbers of nascent transcripts are not (they are obtained, for example, by GRO-seq). It is for this reason that our tests use human data, but testing can be extended to other species as appropriate data becomes available.

**SUPPLEMENTARY FIGURES**

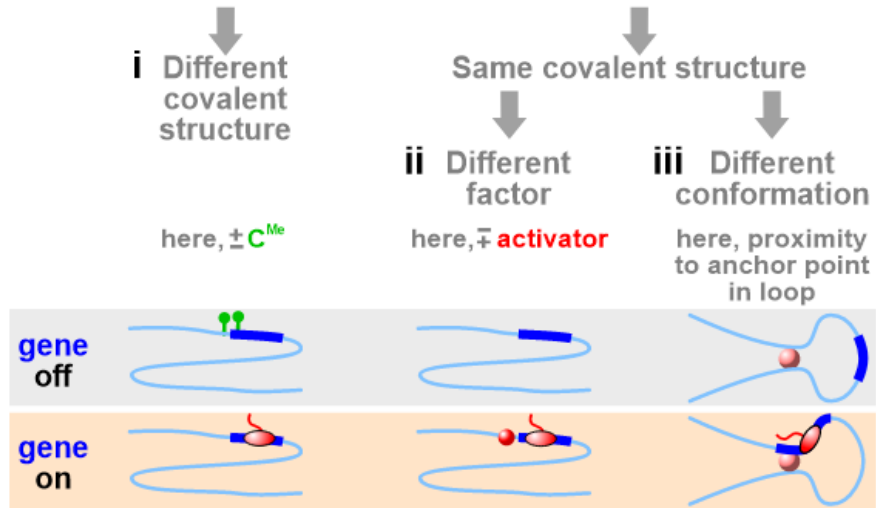

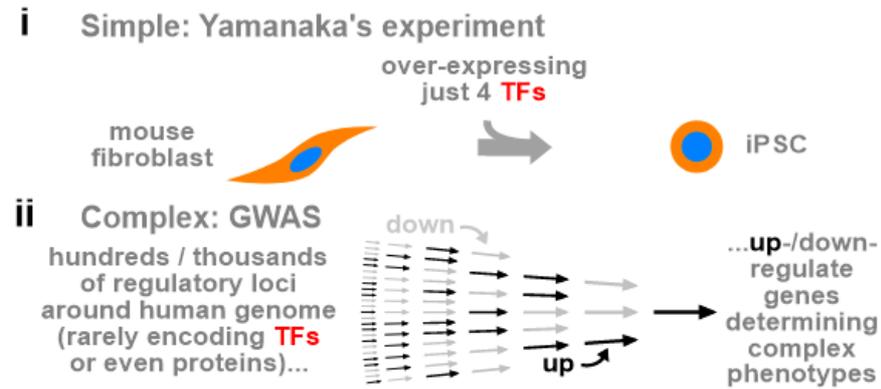

**Supplementary Figure 1.** Ways of switching a gene on/off, and conflicting experimental results.
**A.** Three possibilities. Oval + wavy red line: polymerase + transcript.
(**i**) Covalent DNA structures differ in the off/on states (here, due to methylation of C residues – green lollipops).
(**ii**,**iii**) If covalent structures are identical, one gene may associate with a different factor (here, an activator – red sphere), or adopt a different conformation (here, the polymerase only initiates at the point anchoring a loop).
**B.** Two experimental results give conflicting views of how activity is regulated.
(**i**) Simple (from Yamanaka's experiment).
(**ii**) Complex (from GWAS).

**Supplementary Figure 2.** Major mechanisms stabilizing loops.

**A.** Cohesin is a ring-shaped molecule that clips on to DNA like a carabiner on a climber's rope to stabilize loops (Yatskevich *et al.*, 2019; Davidson & Peters, 2021). [The bridging-induced attraction drives cohesin aggregation (Ryu *et al.*, 2021), which is seen in mouse ES cells – with transcription opposing this (Gu *et al.*, 2020).]

**B.** Phase separation.

**(i)** The catalytic sub-unit of mammalian polymerase II (Boehning *et al.*, 2018; Lu *et al.*, 2018; Plys & Kingston, 2018; Shao *et al.*, 2021), plus initiation (e.g., MED1 with OCT4, ER, MYC, NANOG, SOX2, GATA2) and elongation factors (e.g., CYCT1 of P-TEFb, DYRK1A; Boija *et al.*, 2018) contain low-complexity disordered domains that can form liquid drops.

**(ii)** Such drops can stabilize clusters/loops (Boehning *et al.*, 2018; Boija *et al.*, 2018; Lu *et al.*, 2018; Plys & Kingston, 2018; Shin *et al.*, 2018; Ferrie *at al.*, 2022).

**C.** The depletion attraction (Marenduzzo *et al.*, 2006; Mitchison, 2019).

**(i)** In crowded cells, many small pink molecules (diameter < 5 nm) bombard (grey arrows) larger red complexes (5 – 25 nm) from all sides. If two large complexes collide, smaller molecules are sterically excluded from the green volume between the two and cannot knock them apart; consequently, small molecules exert a force on opposite sides of the two to keep them together.

**(ii)** When polymerases (5-25 nm) bind to DNA this force can drive clustering/looping.

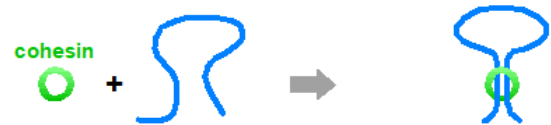
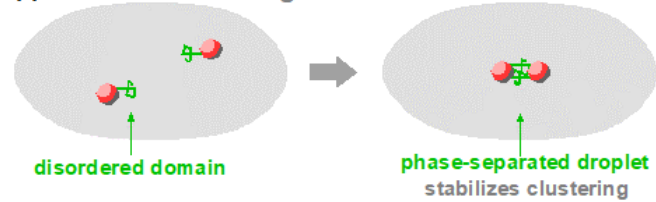
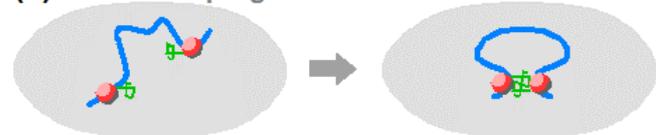
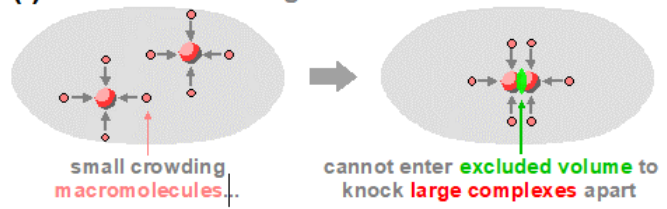
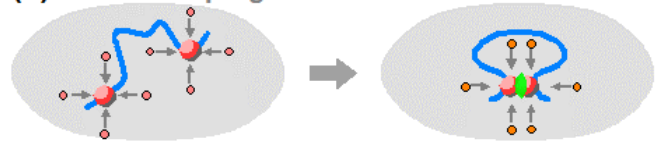

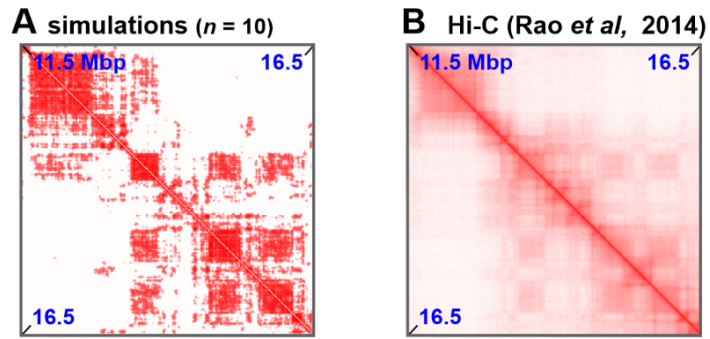

**Supplementary Figure S3.** Simulations involving a simple polymer model yield contact maps like those found experimentally.
**A.** Contact map for a 5 Mbp region of HSA19 in GM12878 cells obtained from 10 simulations of the whole chromosome (data from Brackley *et al.*, 2016).
**B.** Contact map of the same region obtained by Hi-C (data from Rao *et al.* (2014)).

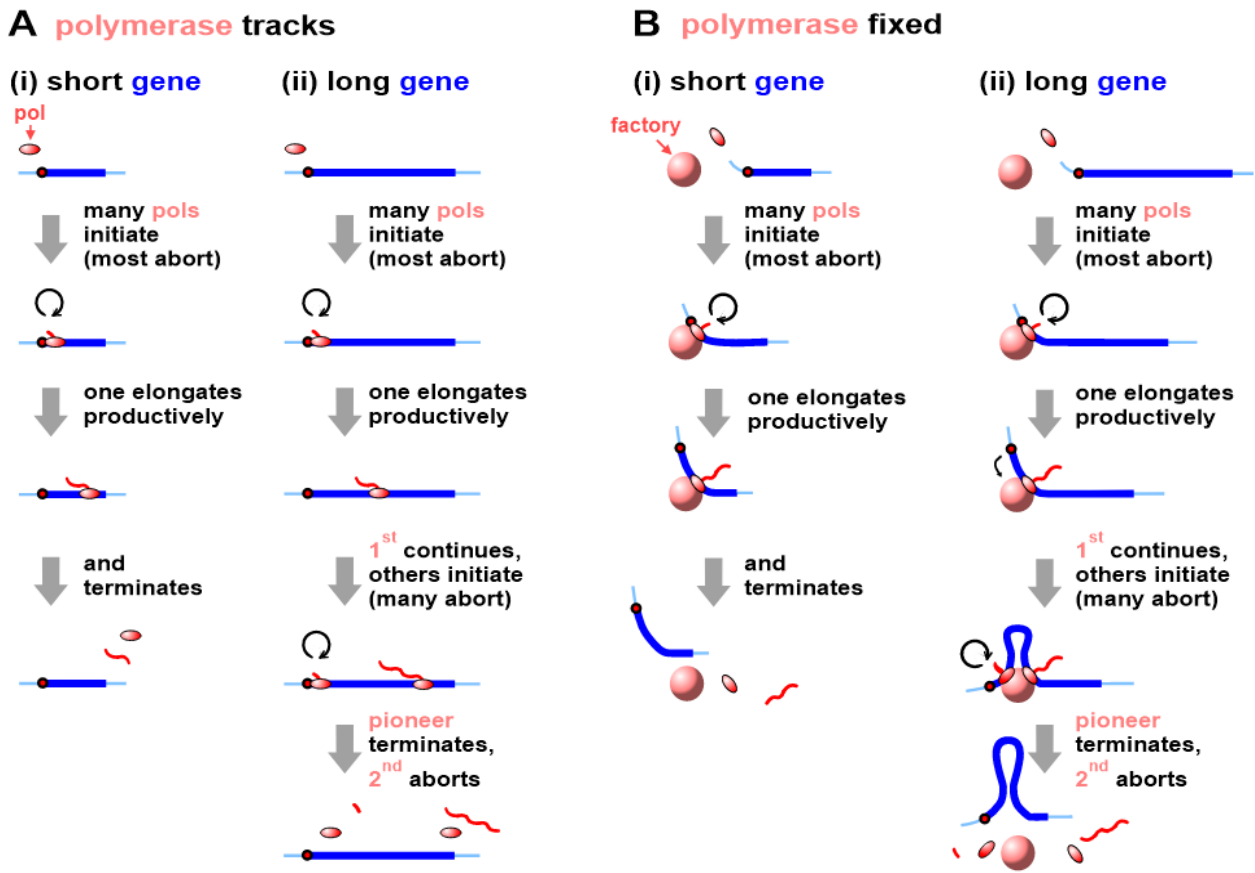

**Supplementary Figure 4.** Transcription cycles (promoter – small red circle; polymerase – red oval; transcript – wavy red line; factory – large pink sphere).

**A.** Traditional model with tracking polymerases.

(**i**) Short gene. Many polymerases bind to a promoter, a few elongate for ~10 nucleotides (then many abort), and fewer still form a productive complex that tracks down the template. A variant model (not shown) has the polymerase initiating in a "hub" and tracking away from the hub (Cramer, 2019).

(**ii**) Long gene. Now a 2$^{nd}$ polymerase may initiate (and probably abort) before the pioneer terminates. Few active human genes are ever loaded with >1 productively-elongating polymerase, and these so-called active genes are idle for most of their time (Larsson *et al*., 2019).

**B.** Alternative with transiently-immobilized polymerases.

(**i**) Short gene. The promoter diffuses to the factory and initiates; most resulting complexes again disassemble. Once a productive complex forms, the polymerase remains transiently bound to the factory as it reels in the template and makes a transcript. As the template moves past the polymerase, close tethering restricts diffusion of the now-active template (as seen; Nagashima *et al*., 2019).

(**ii**) Long gene. Initial steps are as for the short gene, and yield a (productive) elongating complex that tethers the promoter close to the factory (3$^{rd}$ panel down); therefore, the promoter is likely to revisit the same factory (arrow) and reinitiate (as the pioneer continues to transcribe) to yield a "sub-gene loop" (4$^{th}$ panel down). If the 2$^{nd}$ polymerase aborts and the promoter cycles through several initiations and abortions, the length of this sub-gene loop grows as the pioneer continues transcribing. Then, segments lying progressively 3' are brought successively next to the promoter. 3C confirms this with various long human genes (Papantonis *et al*., 2010; Larkin *et al*., 2012; Larkin *et al*., 2013). Super-resolution RNA FISH also confirms that nascent promoter-proximal transcripts lie next to nascent RNAs copied from progressively further into the long gene (Larkin *et al*., 2013). Moreover, "ChIA-Drop" shows such directional 3' bias in contacts between fly promoters and down-stream segments (Zheng *et al*., 2019); in this study ~80% contacts contained one genic promoter – consistent with this model applied to genomes where non-genic promoters out-number genic ones. All these results are simply explained if the active polymerase is transiently immobile, and impossible to explain if the enzyme tracks without complex additional assumptions.

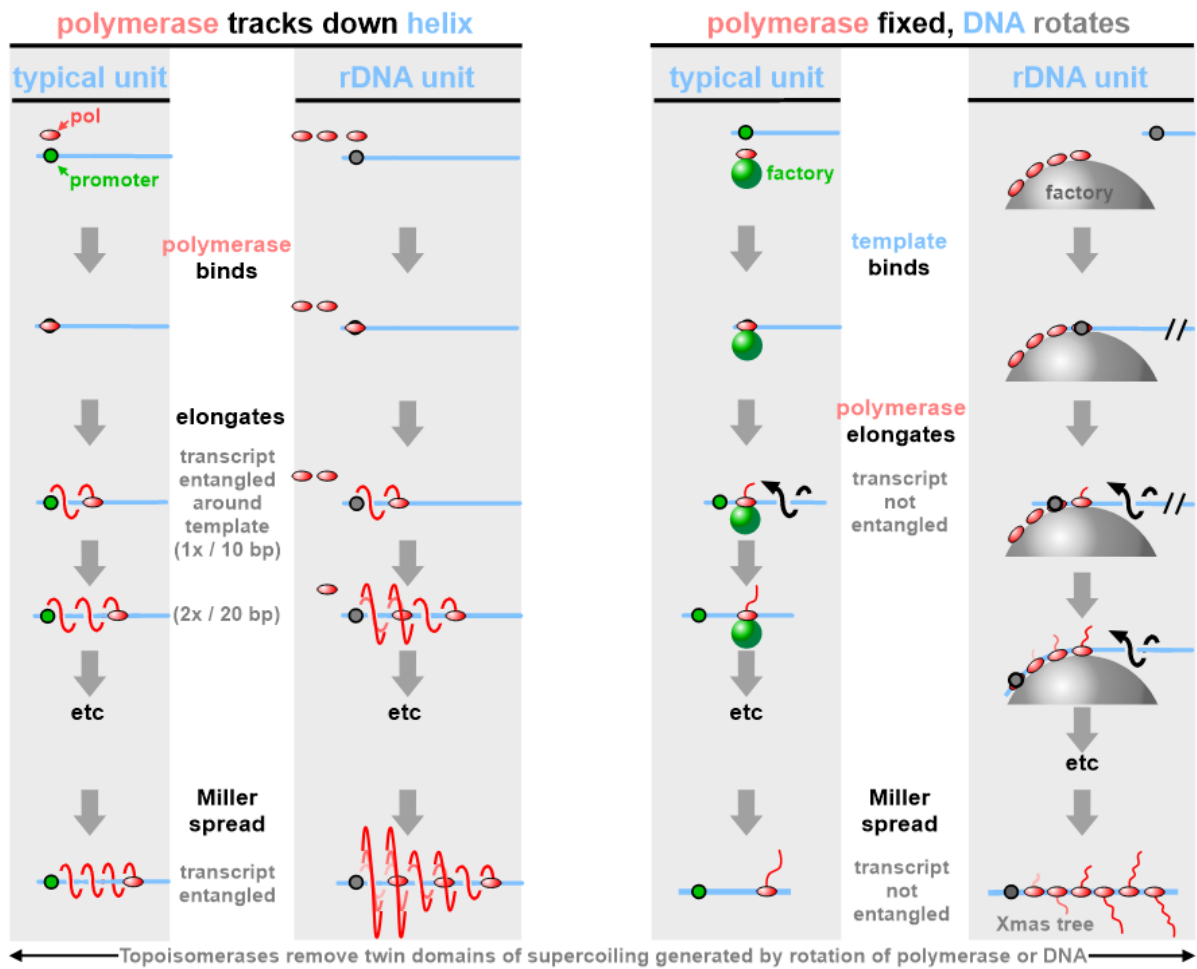

**Supplementary Figure 5.** Interpretation of Miller spreads (Miller *et al.*, 1970; Miller *et al.*, 1972). Iconic images of "Christmas trees" are often cited as the best evidence that active polymerases track down their templates. These are obtained by spreading a 3D structure with >50 polymerases tightly packed on one bacterial rDNA unit on to a 2D surface, and electron microscopy; nascent transcripts appear as "branches" in the tree (as bottom right). Most other TUs are associated with just 1 polymerase and one branch (as bottom of 3$^{rd}$ grey column). We argue such images (and similar ones of lampbrush loops; Papantonis & Cook, 2013) provide strong support for our view and against the traditional one – a view confirmed by single-molecule imaging (Nagashima *et al.*, 2019; Ide *et al.*, 2020). Note that a distinct topological problem arises whether or not polymerases track (twin domains of supercoiling form on each side of the polymerase and are removed by topoisomerases). See also Supplementary Movie 1.

*1$^{st}$ grey column: one polymerase tracking down a typical unit.* Transcription requires lateral and rotational movement along and around a helix, so tracking along a helical path generates a transcript entwined once around a template for every 10 bp transcribed. Spreading an engaged polymerase and 500-nucleotide transcript should yield a transcript with 50 entanglements (only 3 shown at bottom) – and not one extended (un-entangled) "branch" as in Miller's images. No mechanism is known that can untwine a transcript exactly the right number of times to free it from the template to give such images.

*2$^{nd}$ grey column: many polymerases (only 3 shown) tracking down a rDNA unit.* As for a single polymerase, each transcript should become entwined many times. With >50 polymerases we should see a dense mass of RNA around the template (not distinct and untwined "branches" extending from the "trunk").

*3$^{rd}$ grey column: a single enzyme fixed to the green factory reeling in a typical unit.* Now DNA moves laterally plus rotationally (black arrow). Screwing a bolt through a fixed nut provides an analogy. On spreading, the template is stripped off the factory, so we see a single untwined transcript.

*4$^{th}$ grey column: many fixed enzymes (up to 6 shown) reeling in a rDNA unit.* As before, the template moves laterally and rotationally (black arrows) over a larger factory to yield untwined transcripts, and now spreading yields the "Xmas trees" seen in the iconic images of single rDNA units. Here, screwing a bolt through 6 fixed nuts provides the analogy.

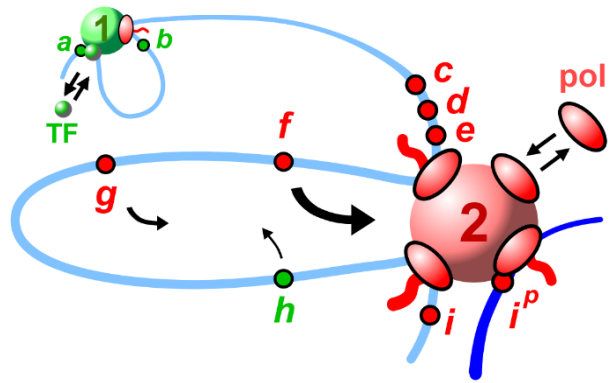

**Supplementary Figure 6.** Regulatory motifs. Letters mark green and red promoters that tend to initiate in numbered factories of similar color. All motifs discussed are transcribed when active.

(**i**) *Enhancers* (Furlong & Levine, 2018; Schoenfelder & Fraser, 2019; Andersson & Sandelin, 2020). Transcription of *e* ensures *f* is tethered close to red 2, so *f* often visits 2 and initiates (creating a new *e:f* loop and 3C contact). Then, we call *e* an enhancer of *f*.

(**ii**) *SEs* (Hnisz *et al.*, 2017). *e* detaches from 2 when its polymerase terminates. If *c*, *d*, and *e* are strong promoters, there is a good chance one re-initiates in 2 to re-tether *f* close to 2 and increase its likelihood of re-initiation. Consequently, we call *c*, *d*, and *e* a SE of *f*. Repeated initiation of *f* generates the bursty transcription seen *in vivo* and in simulations (Brackley *et al.*, 2021).

(**iii**) *Silencers* (Pang *et al.*, 2023). Transcription of *i* ties green *h* close to red 2, and far from green 1. While *h* and *f* visit red 2 equally often, *f* does not initiate there as green factors are missing; moreover, while *i* is transcribed, *h* cannot visit green 1. Consequently, we call *i* a silencer of *h*.

(**iv**) *Boundaries*, *insulators* (Hsieh *et al.*, 2020; Krietenstein *et al.*, 2020). If *b* and *c-e* are transcribed often, *h* visits green 1 rarely; consequently, *b* + *c-e* prevent *f* and *h* from diffusing further afield, and so we call them boundaries and insulators.

(**v**) *eQTLs* and *QTLs*. If a SNP in *e* reduces TF binding, this shortens the time *e* is bound to red 2 – to decrease close tethering (and so firing frequency) of *f* and *g*. If *f* and *g* are genes, we call this SNP an eQTL as it down-regulates both. As *f* and *g* bind red TFs, they are functionally related; this explains why eQTLs target (and often contact) functionally-related genes. Now consider a SNP in *i* that reduces binding to red 2. This shortens the time *i* is bound to 2 and allows *h* to visit green 1 and so fire more often. If *h* is a gene, we will call this SNP an eQTL as it up-regulates *h*. If *f*, *g*, and *h* encode proteins affecting a trait of interest, we call these SNPs in *i* and *h* QTLs. Critically both eQTLs and QTLs act co-transcriptionally here (of course, this does not exclude them from acting post-transcriptionally as in the omnigenic and other models).

(**vi**) Motifs leading to loose chromosome pairing during meiosis and transvection (Xu & Cook, 2008). Homologs have similar DNA sequences and so organize similar strings of colored factories (which we will call "homologous factories") down their lengths. If $i^p$ is the paternal homolog of maternal *i*, homologs are tied together through red factory 2. Binding of more promoters to appropriate homologous factories will then zip chromosomes together. We suggest this occurs rarely in somatic cells, but often in meiotic ones – where complete zipping is aided by long times, plus a variety of species-specific mechanisms (Kim *et al.*, 2022). In human meiosis, we anticipate such a (1st) homology search depends on promoters sifting through a few thousand factories/cell to find appropriately-colored ones. Subsequently, the (2nd) well-known base-pairing search screens through thousands of bases/loop (not the billions in the genome) to achieve tight recombinational pairing (attainable in the time available, even when loops contain repeated sequences that defeat a genome-wide search). Now consider transvection – the *trans* complementation originally seen in flies where mutations on different homologs recreate the wild-type, but only if homologs pair (Fukaya & Levine, 2017). Imagine the maternal homolog encodes wild-type gene $f^m$ plus a mutant enhancer $e^m$ (so $f^m$ is silent), and the paternal one a mutant $f^p$ plus a wild-type $e^p$ (so $f^p$ is also silent). Once homologs pair, wild-type $f^m$ becomes tethered close to a red homologous factory, so enabling it to fire. This factory-based search is aided by local concentrations (in relevant fly cells homologs are polytenized >100x, and in meiosis each pairing partner is duplicated). [It is also driven by the depletion- and bridging-induced attractions that operate throughout evolutionary time, and once bi-/multi-valent proteins appear, respectively. Therefore, both could occur before the Darwinian threshold marking the transition from primordial cells (each probably with many genomes/cell), through LUCA (the last universal common ancestor), to distinct archaeal, bacterial, and eukaryotic lineages (Woese, 2012). This has obvious implications for which paths evolution followed, and when/how the mystery that is sexual dimorphism developed.]

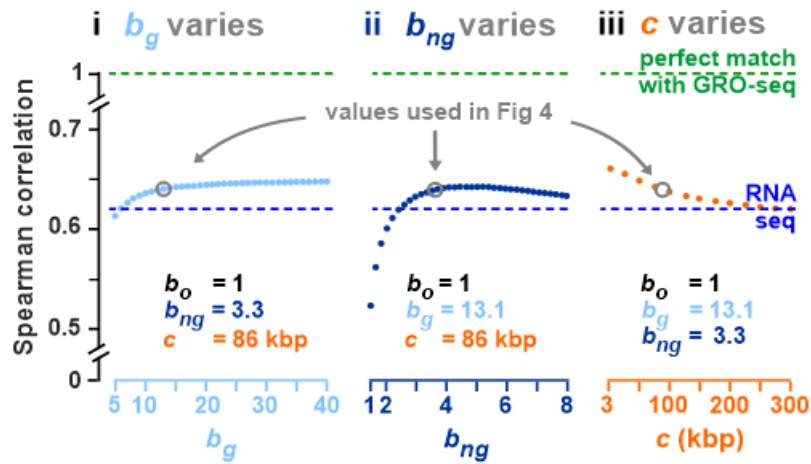

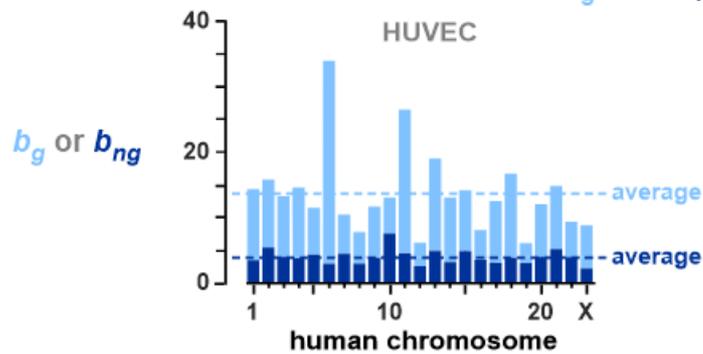

**Supplementary Figure 7.** Artificially varying constants within experimentally-observed limits has little effect on the performance of the 3-state formula.

**A.** Effects of artificially varying of $b_g$, $b_{ng}$, and $c$ (HUVEC, HSA14). The formula is applied as the value of one constant is systematically varied (invariant values indicated); then, rank orders of all promoter activities are determined, and Spearman correlations determined using GRO-seq data as a reference. Grey circles indicate values used in Figure 4. Spearman correlations prove to be relatively insensitive to variations of $b_g$ and $b_{ng}$ (within experimentally-determined limits seen in panel B), and over a wide range of possible average loop lengths.

**B.** Chromosomes-specific values of $b_g$ and $b_{ng}$ determined from GRO-seq data (HUVEC). For each chromosome, promoters are classified as $b_g$, $b_{ng}$, or $b_o$, the number of reads seen in GRO-seq data extracted, values normalized relative to those seen with all $b_o$ promoters (where the average value is set equal to 1), and averages determined. Chromosome-specific values of $b_g$ are always greater than those of $b_{ng}$ (dotted lines give average values). For all chromosomes in the 3 cell types examined (i.e., HUVEC, GM12878, H1-hESC), values of $b_g$ vary between 3-35, and of $b_{ng}$ between 1-8 (not shown).

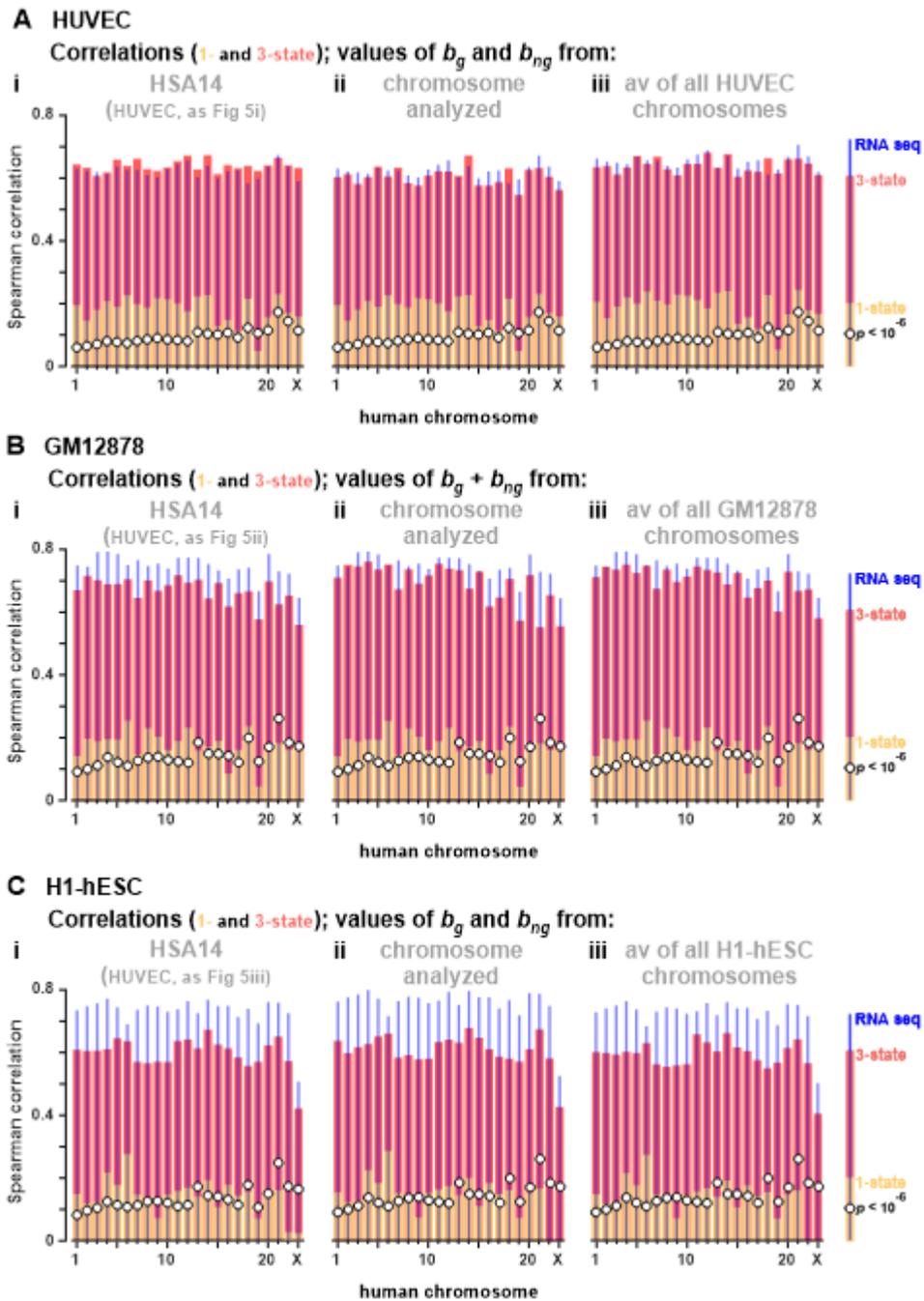

**Supplementary Figure 8.** Varying constants $b_g$ and $b_{ng}$ has little effect on Spearman correlations obtained in different cell types. Predictions obtained with the 1- and 3-state formulae are compared with those obtained experimentally by RNA-seq and GRO-seq for all chromosomes in the cell. With the formulae, active promoters are identified using DHSs, their $p_{trans}$ values calculated and ranked from high to low, and Spearman correlations between the rank order and that from GRO-seq determined (using $c = 86$ kbp). For the 3-state formula, values of $b_g$ and $b_{ng}$ are derived from GRO-seq data using the chromosomes indicated. Note that the p values computed measure the likeliness that correlations are obtained by chance, hence these are identical in all panels in one row. In the 1st panel in each row, values of $b_g$ and $b_{ng}$ are based on those found in HSA14 ($b_g$ = 13.1, $b_{ng}$ = 3.3, and $b_o$ = 1) and are reproduced from Figure 5 to allow comparison. In the 2nd panel, they are for each individual chromosome in that cell, and applied to each chromosome individually. In the 3rd panel, the average value for all chromosomes in that cell are applied to each chromosome individually. Spearman correlations prove to be relatively insensitive to the variations tested.
**A.** Results for HUVEC cells.
**B.** Results for GM12878 cells.
**C.** Results for H1-hESC cells.

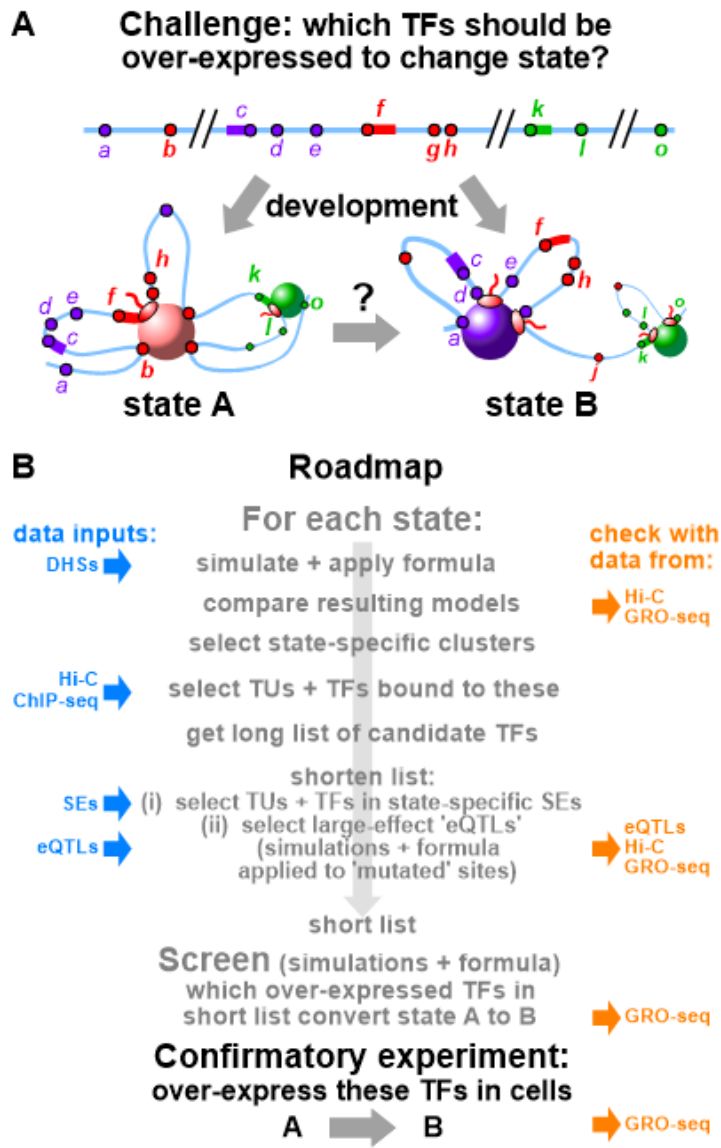

**Supplementary Figure 9.** Solving a grand challenge: the *ab initio* determination of which TFs to over-express to switch any human cell from state A to B. We know TFs specify cell identity by binding to cognate sites associated with unique SEs (Whyte *et al.*, 2013; Boija *et al.*, 2018), that frequent promoter:promoter contacts form sub-maximal cliques (Choy *et al.*, 2018; Liu *et al.*, 2018; Dotson *et al.*, 2022), and that chromosomes in different states have their own sets of characteristic structures (Di Stefano *et al.*, 2020).

**A.** Cartoon: hypothetical genetic map and example structures found in states A + B. TUs *a-o* are in alphabetical order (genic ones – *c*, *f*, *k*; non-genic – all others). Red promoters are active in state A, purple ones in state B, and green (housekeeping) ones in both states. In this conformation of state A, red gene *f* is transcribed (*g* and *h* are its SE – both currently untranscribed), as is housekeeping-gene *k*. In state B, purple gene *c* is tethered close to the purple factory by *d* (which forms a SE with *e*) and will soon fire, and housekeeping *k* is again being transcribed. The grand challenge is to find which TFs to over-express to convert state A to B.

**B.** One possible roadmap. For each state, we identify all active promoters (e.g., using DHS or ATAC-seq data; Zhang *et al.*, 2021), generate sets of 3D structures using simulations (e.g., with HiP-HoP; Buckle *et al.*, 2018) + the formula, check results (e.g., using GRO-seq, NET-seq, Hi-C, micro-C, GAM, pore-C), and select clusters uniquely present in each state. We now identify clusters in the state-specific sub-set rich in particular TFs that specialize in transcribing state-related TUs. Thus, red TUs associated with pink factories in state A will tend to contact other red TUs and so constitute a small-world network detectable by mapping close DNA:DNA contacts (e.g., using Hi-C). For example, *b*, *f*, *g*, and *h* all form frequent pairwise contacts with each other, and with other red TUs on the same and different chromosomes. This small-world of TUs should all bind the same red TFs (detected using ChIP-seq, with contacts between members of the clique being confirmed by Hi-C). This combination will probably yield one long list of red state-specific TUs/TFs, plus another for purple state-specific ones. We now shorten such lists by selecting (i) critical TFs specifying state B that we assume bind to B-specific SEs and promote transcription of B-specific/purple genes, and (ii) large-effect eQTLs affecting B-specific genes (using simulations + the formula, and checking that appropriate contacts are made in each distinct small-world network) by eliminating binding to sites in each SEs in turn to see what consequential effects there are on computed transcriptional activity. Having shortened the list, we screen (using simulations + the formula) to see whether "over-expressing" remaining TFs converts state A to B, and select ones inducing the wanted change. Finally, we confirm experimentally this fate switch. Obviously, the roadmap will be augmented by other approaches (e.g., Cahan *et al*., 2014; Rackham *et al*., 2016; Ronquist *et al*., 2017; Dunn *et al*., 2019; Di Stefano *et al*., 2020; Liu *et al*., 2020; Avsec *et al*., 2021; Kamimoto *et al.,* 2023).

**SUPPLEMENTARY MOVIE 1.** Tracking versus fixed polymerases. The nut, bolt, and string represent a polymerase (pol), template, and transcript, respectively.